\documentclass[prl,twocolumn,english,superscriptaddress]{revtex4-1}

\usepackage[T1]{fontenc}
\usepackage[latin9]{inputenc}
\usepackage{verbatim}
\usepackage{color}
\usepackage{xcolor}
\usepackage{babel}
\usepackage{latexsym}
\usepackage{float}
\usepackage{amsmath}
\usepackage{amsfonts}
\usepackage{graphicx}
\usepackage{times}   
\usepackage{esint}
\usepackage{subfigure}
\usepackage{verbatim}
\usepackage{algorithm}
\usepackage{algorithmicx}
\usepackage{algpseudocode}
\usepackage{soul}

\usepackage[unicode=true,pdfusetitle,
 bookmarks=false,colorlinks=true,citecolor=blue,urlcolor=blue,linkcolor=red]{hyperref}
 
\newcommand{\mc}[2]{\multicolumn{#1}{|c|}{#2}}
\newcommand{\bra}[1]{\left\langle #1 \right|}
\newcommand{\ket}[1]{\left|#1\right\rangle}

\makeatletter

\@ifundefined{textcolor}{}
{%
 \definecolor{BLACK}{gray}{0}
 \definecolor{WHITE}{gray}{1}
 \definecolor{RED}{rgb}{1,0,0}
 \definecolor{GREEN}{rgb}{0,1,0}
 \definecolor{BLUE}{rgb}{0,0,1}
 \definecolor{CYAN}{cmyk}{1,0,0,0}
 \definecolor{MAGENTA}{cmyk}{0,1,0,0}
 \definecolor{YELLOW}{cmyk}{0,0,1,0}
 \definecolor{UCGOLD}{RGB}{183, 135, 39}
}

\@ifundefined{date}{}{\date{}}
\AtBeginDocument{
  
}
\makeatother

\newcommand{\SAVE}[1]{}

\begin{document}
\renewcommand\abstractname{}

\title{Postponing the orthogonality catastrophe: efficient state preparation for electronic structure simulations on quantum devices}
\author{ Norm M. Tubman}
\affiliation{Kenneth S. Pitzer Center for Theoretical Chemistry, Department of Chemistry, University of California, Berkeley, California 94720, USA and Chemical Sciences Division, Lawrence Berkeley National Laboratory Berkeley, California 94720, USA}
\author{Carlos Mejuto-Zaera}
\affiliation{Kenneth S. Pitzer Center for Theoretical Chemistry, Department of Chemistry, University of California, Berkeley, California 94720, USA and Chemical Sciences Division, Lawrence Berkeley National Laboratory Berkeley, California 94720, USA}
\author{Jeffrey M. Epstein}
\affiliation{Department of Physics, University of California, Berkeley, California 94720, USA}
\author{Diptarka Hait}
\affiliation{Kenneth S. Pitzer Center for Theoretical Chemistry, Department of Chemistry, University of California, Berkeley, California 94720, USA and Chemical Sciences Division, Lawrence Berkeley National Laboratory Berkeley, California 94720, USA}
\author{Daniel S. Levine}
\affiliation{Kenneth S. Pitzer Center for Theoretical Chemistry, Department of Chemistry, University of California, Berkeley, California 94720, USA and Chemical Sciences Division, Lawrence Berkeley National Laboratory Berkeley, California 94720, USA}
\author{William Huggins}
\affiliation{Kenneth S. Pitzer Center for Theoretical Chemistry, Department of Chemistry, University of California, Berkeley, California 94720, USA and Chemical Sciences Division, Lawrence Berkeley National Laboratory Berkeley, California 94720, USA}
\author{Zhang Jiang}
\affiliation{Google Inc., Venice, California 90291, USA}
\author{Jarrod R. McClean}
\affiliation{Google Inc., Venice, California 90291, USA}
\author{Ryan Babbush}
\affiliation{Google Inc., Venice, California 90291, USA}
\author{Martin Head-Gordon}
\affiliation{Kenneth S. Pitzer Center for Theoretical Chemistry, Department of Chemistry, University of California, Berkeley, California 94720, USA and Chemical Sciences Division, Lawrence Berkeley National Laboratory Berkeley, California 94720, USA}
\author{K. Birgitta Whaley}
\affiliation{Kenneth S. Pitzer Center for Theoretical Chemistry, Department of Chemistry, University of California, Berkeley, California 94720, USA and Chemical Sciences Division, Lawrence Berkeley National Laboratory Berkeley, California 94720, USA}
\date{\today}
\begin{abstract} 

Despite significant work on resource estimation for quantum simulation of electronic systems,
the challenge of preparing states with sufficient ground state support has so far been largely neglected.  In this work we investigate this issue in several systems of interest, including
organic molecules, transition metal complexes, the uniform electron gas, Hubbard models, and quantum impurity models arising from embedding formalisms such as dynamical mean-field theory. Our approach uses a state-of-the-art classical technique for high-fidelity ground state approximation. We find that easy-to-prepare single Slater determinants such as the Hartree-Fock state often have surprisingly robust support on the ground state for many applications of interest. For the most difficult systems, single-determinant reference states may be insufficient, but low-complexity reference states may suffice. For this we introduce a method for preparation of multi-determinant states on quantum computers.

\end{abstract}
\maketitle

\newpage


\newcommand{\proj}[1]{\ket{#1}\hspace{-0.4pt} \bra{#1}}
\newcommand{\ndet}{L}
\newcommand{\norb}{n}
\newcommand{\scnot}{\textsc{not}}
\newcommand{\DD}{D}
\newcommand{\pp}{p}
\newcommand{\una}{\mathrm{una}}
\newcommand{\sys}{\mathrm{sys}}
\newcommand{\ins}{\mathrm{in}}
\newcommand{\dd}{d}

\section*{Introduction}
\label{sec:intro}
Simulation and characterization of systems of interacting fermions has long motivated the development of quantum computers ~\cite{
Abrams1997,Ortiz2001}. In particular, characterization of ground state properties of chemical systems \cite{guzik2005} has been proposed as an area likely to be impacted by near-term quantum computers~\cite{mcclean2016,babbush2018,peruzzo2014,kivlichan2018quantum}. 
However, establishing a quantum advantage in this domain is challenging, given the decades of work on a large number of classical tools for treating quantum systems.  
Nonetheless, it seems clear that quantum algorithms will offer solutions to difficult problems in electronic structure that fall beyond the reach of current classical techniques. Specifically, accurate quantum simulation of long-time dynamics and determination of spectra \cite{Kitaev1995,Abrams1999} are expected to be powerful tools superior to any classical approach ~\cite{Jones2012,wecker2014,Reiher2017,BabbushSpectra}. 
Indeed, experiments on several quantum computing architectures have already implemented the phase estimation algorithm to determine low-lying energies of small molecules \cite{lanyon2010towards,Wang2014,OMalley2016}.

As quantum technologies advance, 
increasing efforts have sought to estimate the resources that will be required to solve classically intractable electronic structure problems \cite{Whitfield2010,wecker2014,Reiher2017,BabbushSpectra}. However, resource estimates thus far have ignored the problem of initial state preparation for phase estimation, largely due to the lack of classical tools capable of addressing this issue. 
Concern with the fidelity of a ground state ansatz, as determined by the overlap between an approximate and true wavefunction of a physical system, has a long history. One can make a simple prediction that the overlap of an approximate wavefunction with the true wavefunction will fall off exponentially as a function of system size, even if local observables remain relatively consistent. This phenomenon was referred to in Walter Kohn's Nobel Prize speech as the ``Van Vleck Catastrophe''~\cite{Kohn1999nobel} and is also known as the orthogonality catastrophe.  While this leads some to worry 
about the interpretation of wavefunctions for macroscopic systems, in the context of the phase estimation algorithm outlined above, it has clear practical implications. 

The catastrophe is a critical aspect of resource accounting for quantum phase estimation, because success of the algorithm requires significant overlap with the ground state.  
For example, recent resource estimates have suggested that phase estimation on FeMoco, a transition metal center used for splitting the nitrogen dimer in the nitrogenase enzyme~\cite{Reiher2017}, will be possible with near-term quantum computers given a good ground state estimate as a starting point, yet it is unclear that finding such a starting wave function is feasible.  
Due to the expected strong correlations in FeMoco, even variational quantum eigensolvers (using, for example, a unitary coupled cluster wavefunction ansatz) ~\cite{Bartlett1989,Romero2017} or adiabatic state preparation ~\cite{Wu2002,BabbushAQChem} approaches will most likely not be sufficient for this task.
Other recent work \cite{BabbushSpectra} has  performed resource counts for simulation of solid-state electronic structure systems, including diamond, graphene, lithium metal, and the uniform electron gas, but has also neglected the ground state overlap issue. 

In this paper, we contribute to the understanding and practical application of phase estimation applied to large and strongly-correlated systems by addressing two related but distinct questions:
\begin{enumerate}
\item Do there exist efficiently-findable and preparable initial states for phase estimation on large systems, i.e., is phase estimation likely to be efficient when the cost of state preparation is included in the analysis?

\item If there are such states, how do we prepare them, particularly in the case that they are moderately complex compared to product states and/or single-determinant states?
\end{enumerate}
We investigate a range of chemical and solid-state systems, demonstrating that in many cases, 
state preparation at the mean-field level is not only efficient but also yields high-quality initial states for phase estimation, even for fairly large systems with strong correlations.  
This work goes beyond previous estimates of relevant state overlaps ~\cite{Ward2009,wecker2014,McClean2014} by using a state-of-the-art classical method, the adaptive sampling configuration interaction method (ASCI)~\cite{tubman2016-1}.  Combined with a new method for preparing multi-determinant states on quantum registers, this application of ASCI  facilitates quantum computational treatment of systems consisting of up to hundreds of orbitals, far larger than previously considered with classical algorithms and approaching sizes where novel chemical and solid state insights will become possible.

\section*{Adapative Sampling Configuration interaction method (ASCI)}
The main idea behind the ASCI approach is to perform diagonalization on a determinant space in which one captures as many important degrees of freedom as possible. While this is the principle behind all exact diagonalization and CI techniques, most methods do not allow for explicit searching for important determinants~\cite{lauchli2011,vogiatzis2017,szabo:book,gan2006,gan2005,sherrill1999,abrams2004,szalay2012,bender1969,buenker1978,roth2009,bagus1991}. 
In contrast with the more traditional CI techniques, the idea of using a selected CI approach is to generate a relatively small set of determinants that account for 90\% or more of the top contributions to the full CI wavefunction~\cite{tubman2016-1}.

As in most selected CI methods, ASCI proceeds by iteratively improving a wavefunction $\psi_{k}$ to reach a desired accuracy. The search component of the algorithm requires two rules, a selection criterion to determine what part of Hilbert space to search for new determinants (pruning), and a ranking criterion to determine the best determinants to include in the improved wavefunction $\psi_{k+1}$. 

For the ASCI algorithm, the ranking criterion is derived from a consistency relation among the coefficients of determinants in the ground state. Expressing the time-independent Schr\"odinger equation $H\ket{\psi}=E\ket{\psi}$ in the basis of determinants $\{\ket{D_i}\}$, so that $\ket{\psi}=\sum_iC_i\ket{D_i}$, 
we can rearrange to obtain
\begin{equation}
C_{i} = \frac{\sum_{j \ne i}H_{ij}C_{j}}{E-H_{ii}},
\end{equation} 
where $H_{ij}$ is the Hamiltonian matrix element between the $i^\text{th}$ and $j^\text{th}$ determinants.  This equation can be reinterpreted to generate an improved set of determinants by taking the left-hand side as an estimate of the magnitude of the expansion coefficients $C_i^{k+1}$ and replacing $E$ by the energy of the wavefunction in the $k^\text{th}$ iteration. We rank the determinants by magnitude of $C_i^{k+1}$. These coefficients are related to a first-order perturbation estimate for CI coefficients in many body perturbation theory~\cite{huron1973}. 

In practice, this iterative approach generates all the top contributions to the ground state wavefunction. Having the top contributions is critical to obtain highly accurate energies, as was recently shown with the ASCI method in combination with second order many body perturbation theory~\cite{tubman2016-1}.

\section*{Preparing multi-determinant states}

Many interesting chemical and solid-state systems have strongly-correlated ground states characterized by nearly zero overlap with any product wave function of single-electron orbitals.
As a result of this low overlap, such product states, while easy to prepare on a quantum computer, are of limited use for  application of phase estimation. While various mean-field states have been proposed to solve this problem in different instances (see e.g. \cite{wecker2015solving}), the complexity of the true ground state wave functions is such that these states are unlikely to provide a general solution.  It will therefore be necessary to prepare initial states with sufficient complexity, a challenge that we investigate in detail in later sections.
In the rest of this section, we first describe methods for preparation of arbitrary superpositions of multiple Slater determinants on a quantum computer, filling a gap in the phase estimation toolkit. 

The aim is to prepare a quantum state
\begin{align}\label{eq:psi_in}
 \ket{\psi_\ins} = \sum_{\ell=1}^\ndet \alpha_\ell \ket{\DD_\ell}\,
\end{align}
with $\ket{\DD_\ell}$ is a computational basis state corresponding to a bit-string encoding the occupation of the orbitals. We assume that the number of determinants is much smaller than the number of $n$-qubit states, $\ndet \ll 2^\norb$. 

One way to prepare this state is by introducing a ``compressed'' register of $\log \ndet$ qubits whose basis states $\ket{\ell}$ correspond to the determinants $\ket{\DD_\ell}$. We prepare a state of the compressed register, 
\begin{align}
 \ket{\phi} =  \sum_{\ell=1}^\ndet \alpha_\ell \ket{\ell}\,,
\end{align}
which may be achieved with $\mathcal O (\ndet)$ gates \cite{Shende_synthesis_2006}, and implement an isometry to map the basis states of the compressed register to the corresponding encoded Slater determinants on $\norb$ qubits,
\begin{align}\label{eq:l_to_Dl}
 \ket{\ell} \mapsto \ket{\DD_\ell},\quad \text{for all $\ell=1,2,\dots,\ndet$}. 
\end{align}
This transformation can be implemented by the select unitary method introduced by Childs \emph{et al.}~\cite{Childs_Maslov_Nam_Ross_Su_2017}, or equivalently and more generally by the quantum read-only memory (QROM) method introduced by Babbush \emph{et al.}~\cite{BabbushSpectra}.

This method may be improved upon by eliminating the need for a compressed register and instead using a single auxiliary qubit. Assume that we have already prepared the $\norb+1$ qubit state
\begin{align}\label{eq:iter_state}
 \ket{\psi_\ell} = \beta_\ell \ket{\DD_\ell}\ket{1} + \sum_{\ell' = 1}^{\ell-1} \alpha_{\ell'} \ket{\DD_{\ell'}}\ket{0}\,,
\end{align}
where the $\alpha_{\ell'}$ are the coefficients in the target state and $\lvert\beta_\ell\rvert$ may be derived from normalization. Picking any qubit $k$ whose value differs in $\ket{\DD_\ell}$ and $\ket{\DD_{\ell+1}}$, we apply a rotation on that qubit controlled by the auxiliary register:
\begin{align}\label{eq:2d_rotation_a}
 \beta_\ell \ket{\DD_\ell}\ket{1} \mapsto \big(\alpha_\ell \ket{\DD_\ell} + \beta_{\ell+1} X_k \ket{\DD_\ell}\big)\ket{1}\,.
\end{align}
We then erase the auxiliary register value for $\ket{\DD_\ell}$,
\begin{align}\label{eq:2d_rotation_b}
 \ket{\DD_\ell}\ket{1} \mapsto \ket{\DD_\ell}\ket{0}\,,
\end{align}
which takes $\mathcal O(\norb)$ gates to implement. At this point, only the branch of the wave function in which we are trying to construct $\ket{D_{\ell+1}}$ has the auxiliary qubit in the state $\ket{1}$. We then implement the \scnot\ gate, controlled by the unary register, on the remaining qubits where $\ket{\DD_\ell}$ and $\ket{\DD_{\ell+1}}$ are different. This completes the transformation
\begin{align}\label{eq:2d_rotation_c}
 \beta_\ell \ket{\DD_\ell}\ket{1} \mapsto \alpha_\ell \ket{\DD_\ell}\ket{0} + \beta_{\ell+1} \ket{\DD_{\ell+1}}\ket{1}\,.
\end{align}
Combining Eqs.~(\ref{eq:2d_rotation_a})-(\ref{eq:2d_rotation_c}), we have
\begin{align}
 \ket{\psi_\ell} \mapsto  \ket{\psi_{\ell+1}}\,.
\end{align}
Preparing $\ket{\psi_1}$ is trivial, and we can then proceed inductively. The entire state preparation protocol therefore requires $\mathcal O(\norb \ndet)$ gates to implement. Moreover, the total number of gates can be reduced by ordering the determinants such that the Hamming distances between neighboring determinants are small.

\section*{State Preparation and Overlap Estimation}
\label{sec:estimate}

Equipped with ASCI for high-fidelity determination of ground state wave functions and the new quantum algorithms for multi-determinant state preparation, we now analyze several approaches to state preparation for phase estimation. 
One type of approach uses Hartree-Fock orbitals, constructed to provide the minimum energy single determinant over variation of molecular orbitals. Another uses natural orbitals, approximately obtained by performing ASCI to yield a high-fidelity ground state wave function and diagonalizing the corresponding one particle reduced density matrix (1-RDM). Whereas the former is designed to generate states that minimize energy, the latter in some cases improves overlap with the ground state wave function. In both cases, we consider single- and multi-determinant states, using ASCI to select the most important determinants in the ground state wave function.

In order to evaluate the quality of these state preparation methods, we again make use of ASCI to estimate the overlap of the ansatz with the true ground state. Because we have access to the ground state only insofar as ASCI provides a good approximation, this appears at first glance to introduce a circularity in our approach, whereby we attempt to benchmark a ground state ansatz for use in a better-than-classical algorithm for ground energy estimation using a classical algorithm for ground states. However, this is not the case. To understand why our use of ASCI is valid, it is important to understand the accuracy we are looking for in terms of both energy and overlap.

\begin{figure}[htb!]
\includegraphics[width=\columnwidth]{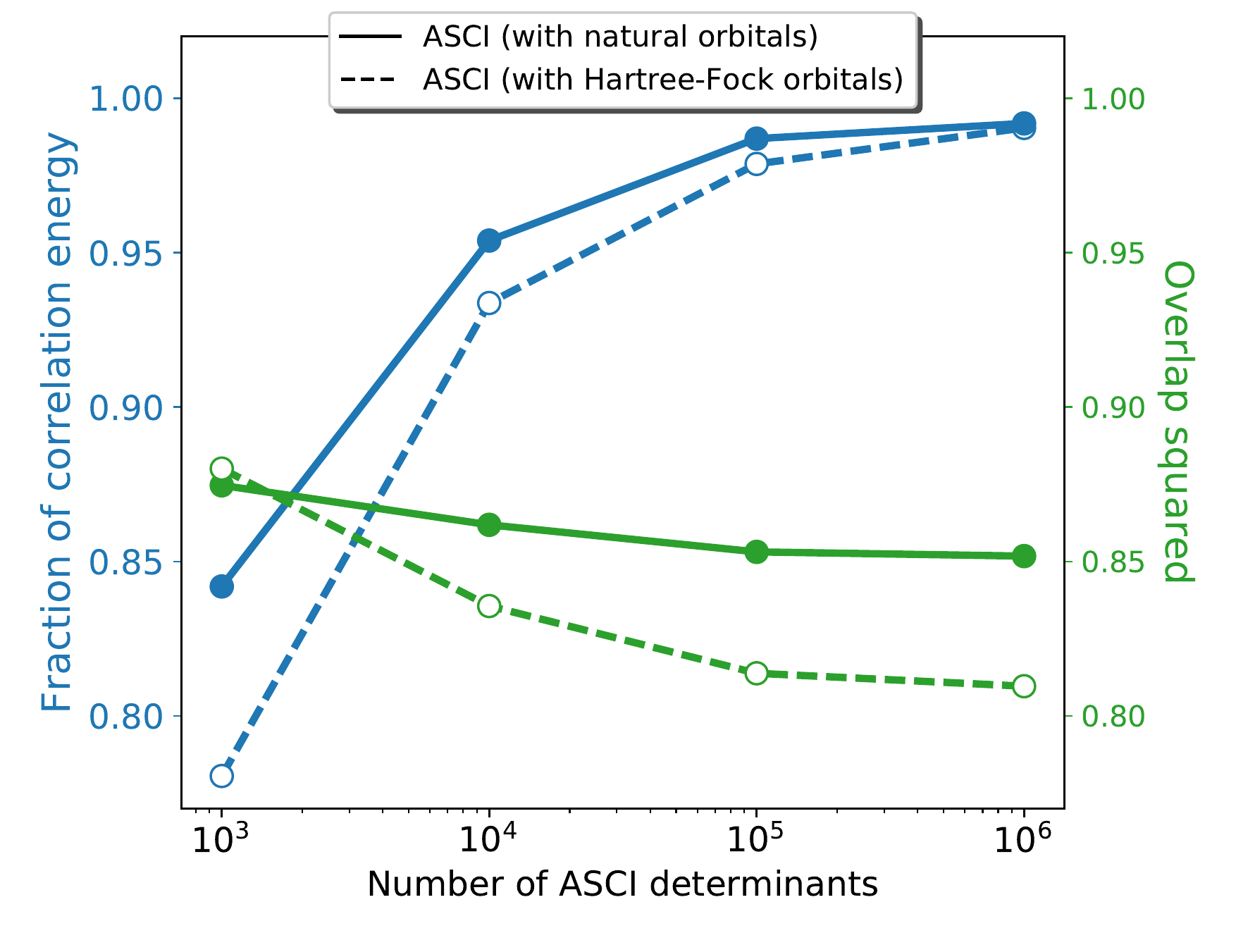}
\caption{Relative convergence rates of ASCI correlation energies, and the square of the overlap of the ASCI wavefunction with the most important Slater determinant for
the CN radical (1.17 {\AA} internuclear separation).
The overlap converges much more quickly than the energy, indicating the possibility of getting good overlap estimates classically without requiring complete energy convergence. It can also be seen that larger overlaps may be obtained via rotating to the natural orbital basis.}
\label{fig:convergence}
\end{figure}

As an illustrative example, in Fig.~\ref{fig:convergence} we demonstrate the convergence of the coefficient of the most important determinant yielded by ASCI for the CN molecule, as a function of the number of determinants included in the wave function.  For molecular systems, we generally look to attain chemical accuracy (1 kcal/mol), which would require more than 99\% of the correlation energy for this molecule.  With regard to overlap for phase estimation, we only want to know if an overlap will be large enough so that we can expect to see the ground state within the number of iterations we can afford to run. For CN, our numerical evidence suggests that the overlaps are well-converged with respect to this criterion well before the correlation energy converges.

\begin{figure}[htb!]
\includegraphics[width=\columnwidth]{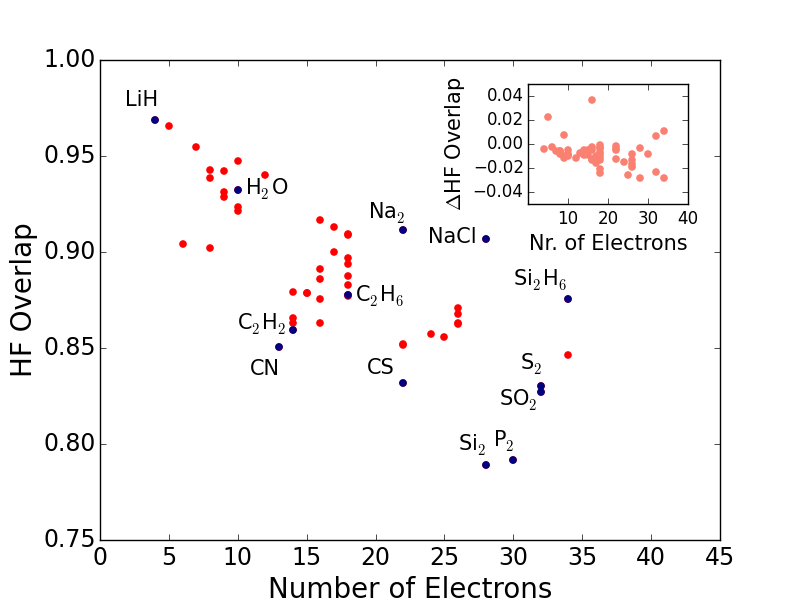}
\\\caption{Scatter plot of systems with respect to number of electrons for cc-pVTZ basis set.  Labeled calculations are highlighted in dark blue.  The inset shows the difference between the Hartree-Fock overlap in the cc-pVDZ and cc-pVTZ basis. See text for more explanation.}
\label{fig:TZvsDZ}
\end{figure}


\section*{Survey of Problems in Electronic Structure}
In this section we explore several paradigmatic chemical systems that will be of interest for early-stage quantum computation. As a benchmark, we examine the G1 set, a classically well-characterized set of small molecules ~\cite{pople1989,tubman2018-1,tubman2018-2,feller1,mardirossian2017thirty,hait2018accurate}. Even these molecules, for realistic basis sets, have large enough Hilbert spaces to be intractable for exact diagonalization. They thus provide valuable examples for exploring the utility of quantum algorithms for many-body systems. We then explore molecules with stretched bonds, which often present difficulties for classical methods as a consequence of their strongly multi-reference nature \cite{gwaltney2002perturbative,van2000quadratic}. Finally, we consider two molecules with transition metal centers:  Iron(II) porphyrin and the FeMo cofactor (FeMoco). Porphyrins are naturally-occurring conjugated organic molecules often complexed with metals in biological systems. Iron(II) porphyrin is part of the heme center in hemoglobin and the hydrocarbon oxidation active site of Cytochrome p450 oxidase enzymes. FeMoco is a cofactor in nitrogenase enzymes that carries out the challenging catalytic reduction of the N$_2$ triple bond to ammonia at ambient conditions, a poorly-understood process critical for agricultural production ~\cite{Reiher2017,sharma2017}. 

We also investigate several systems relevant to condensed matter physics. The Hubbard model is the simplest model for localized electrons in a solid, yet possesses a rich phase diagram \cite{Gull2010,Sordi2012} with ground states featuring radically different types of strongly-correlated order, e.g. anti-ferromagnetic Mott insulation and d-wave superconductivity \cite{Zheng2017}. Dynamical mean-field theory (DMFT), an embedding method for solid-state Hamiltonians \cite{Acharya2018,Hwang2018}, also requires the ability to solve strongly-correlated Hamiltonians \cite{Bauer2016,Gull2010,Sakai2009,Zheng2017,Ehlers2017}. The implementation proposed in \cite{Bauer2016} first prepares the ground state of the impurity model via adiabatic state evolution and quantum phase estimation. Finally, the homogeneous electron gas (HEG) is a Fermi liquid system often used as a baseline for density functional development \cite{ceperley1980,VWN,PW92} and is of intrinsic theoretical interest~\cite{kivlichan2018quantum,shepherd2012investigation,bernu2011,brown2013,malone2016accurate}. Moreover, by changing the Wigner-Seitz radius $r_s$ (a measure of typical electron-electron separation) the correlation strength of the HEG may be arbitrarily tuned, allowing exploration of both strong and weak correlation limits.

\section*{Results}
	\label{sec:theory}

We use the cc-pVTZ basis set of Gaussian spatial orbitals to discretize the Hamiltonians of the G1 set, and find that the corresponding single-determinant Hartree-Fock state is in all cases a good ground state ansatz for the purposes of phase estimation, with all squared overlaps greater than $75\%$, although showing a decrease for increasing number of electrons (Fig. \ref{fig:TZvsDZ}). 
These results appear robust to changes in basis set, as shown in the inset. 
The use of natural orbitals does not offer any substantial improvement in ground state overlap. We conclude that for most molecular systems of interest, the Hartree-Fock state will be a sufficient initial state for phase estimation.

\begin{figure}[htb!]
\begin{center}
\scalebox{1}{\includegraphics[width=\columnwidth]{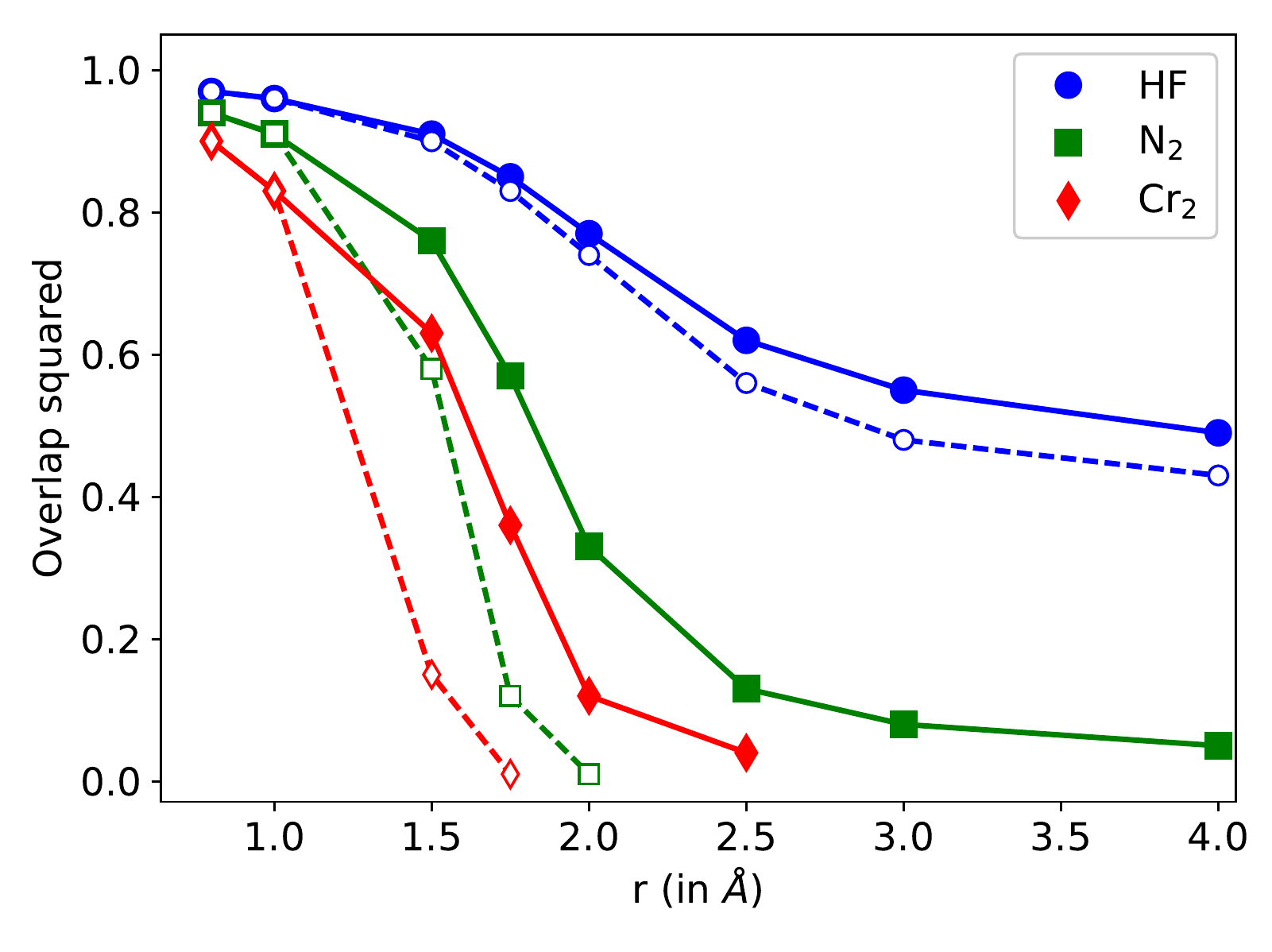}}
\end{center}
\caption{Square of the overlap between the ground state wavefunction and the Slater determinant with aufbau filling for various internuclear separations of HF, N$_2$ and Cr$_2$. All calculations were done with the SVP basis. The solid markers represent calculations where natural orbital rotations were performed in order to have an improved set of orbitals relative to the RHF solutions. The hollow markers denote calculations where the RHF reference was employed directly (i.e. no orbital rotations). It is therefore evident that the overlap decays much more quickly when the naive RHF reference is used, instead of a more optimal natural orbital reference. Equilibrium bond distances are 0.9 {\AA} (HF), 1.1 {\AA} (N$_2$) and 1.68 {\AA} (Cr$_2$) respectively.}
\label{fig:stretches}
\end{figure}

In Fig. \ref{fig:stretches}, we show results for stretched HF, N$_2$, and Cr$_2$. Close to their equilibrium lengths, these molecules show substantial overlap between the true ground state and the single-determinant Hartree-Fock state. For HF, the squared overlap remains large at all separations, approaching $0.5$ at large distances. This is consistent with the expectation that only two determinants play a major role in a stretched single bond.

For large separations of N$_2$ and Cr$_2$ (i.e., large compared to their equilibrium bond lengths), multi-determinant initial states are necessary for good overlaps. The ground state wave function of $N_2$ at $r=4$ {\AA} is reconstructed to a squared overlap of 96\% with 20 determinants over natural orbitals. Cr$_2$ presents qualitatively the same picture, although the overlap decay rate with atomic separation is substantially larger. The ground state wave function at $r=2.5$ {\AA} is reconstructed to a squared overlap of 84\% with 70 determinants over natural orbitals. It therefore appears that good initial states for phase estimation on even very strongly correlated chemical systems can be prepared efficiently on a quantum computer via our multi-determinant state preparation algorithm.

In Table.~\ref{tab:fep} we present overlaps of the ground states of Iron Porphyrin and FeMoco with single-determinant states constructed from Hartree-Fock and natural orbitals for two different active spaces: (32e,29o) and (44e,44o) as discussed in Refs. ~\cite{Reiher2017,sharma2017}. 
In all cases, regardless of spin state, we find squared overlaps between 70-80\%.  
It is often standard to run a CASSCF\cite{roos1980complete} calculation to optimize the orbitals further.  A CASSCF calculation will use an approximate many body wave function to optimize the active orbitals with rotations on inactive orbitals.  Our initial tests with the recently developed  ASCI-SCF approach~\cite{levine2018}, a method for performing a CASSCF optimization with an ASCI wave function, suggest that the overlaps do not change significantly for the iron(II) porphyrin molecule.

\begin{table}
\centering
\begin{tabular}{|c|c|c|c|c|c|c|c|}
\hline
  &\mc{4}{ Iron(II) Porphyrin}  &\mc{1}{FeMoco}     \\\hline
  &\mc{2}{ (32e,29o)}  & \mc{2}{(44e,44o)} &\mc{1}{(54e,54o)}     \\\hline
Orbitals &triplet & quintuplet & triplet & quintuplet& singlet\\\hline
 canonical & 0.81 &	0.82 &	0.73 &	0.76&  0.77\\
 natural 	& 0.82  &	0.82 &	0.78 &	0.78 &0.76 \\
\hline
\end{tabular}
\caption{Square of the overlaps between the selected CI wavefunction and the most important Slater determinant for Iron Porphyrin with and without orbital rotations.  Additional results for FeMoco included with and without orbital rotations. 
}
\label{tab:fep}
\end{table}

In Table \ref{tab:ueg}, we report the overlap squared between the ground state of the HEG and the single-determinant Hartree-Fock state in two and three dimensions for $0.5\le r_s\le 5$, using spaces of (14e, 57o) for 3D and (10e,69o) and (26e,161o) for 2D.
No significant variation in overlap was observed on increasing basis size in either 2D or 3D, indicating that the results are not particularly basis set sensitive. However, increasing the number of electrons is expected to reduce the overlaps for fixed $r_s$, as observed on comparison of the 2D calculations with different numbers of electrons. \label{jar5} 

For fixed $r_s$, the 2D gas features smaller overlaps than the 3D gas, consistent with the observation that low dimensional systems tend to feature strong correlations. In general, overlaps are large up to $r_s=2$, indicating that state preparation for these systems is likely to be easy. For larger values of $r_s$, multi-determinant initial states may be necessary.

\begin{table}[htb!]
\centering
\begin{tabular}{l|l|l|l}
$r_s$  & 2D (10 e$^{-}$)& 2D (26 e$^{-}$)                          & 3D (14 e$^{-}$)                        \\\hline
0.5 & 0.91  & 0.75                      & 0.97                       \\
1   & 0.76 & 0.57                       & 0.90                        \\
2   & 0.53       &0.44                 & 0.75                       \\
5   & 0.22$\pm$ 0.02 & - & 0.41$\pm$0.05\\
10 & 0.15 $\pm$0.03 & - & -
\end{tabular}
\caption{Square of the overlaps between the Fermi gas Hartree-Fock state and the ground state of the homogeneous electron gas in two and three dimensions.}
\label{tab:ueg}
\end{table}



\begin{figure*}
\begin{center}
\scalebox{1}{\includegraphics[width=2.\columnwidth, height=1.\columnwidth]{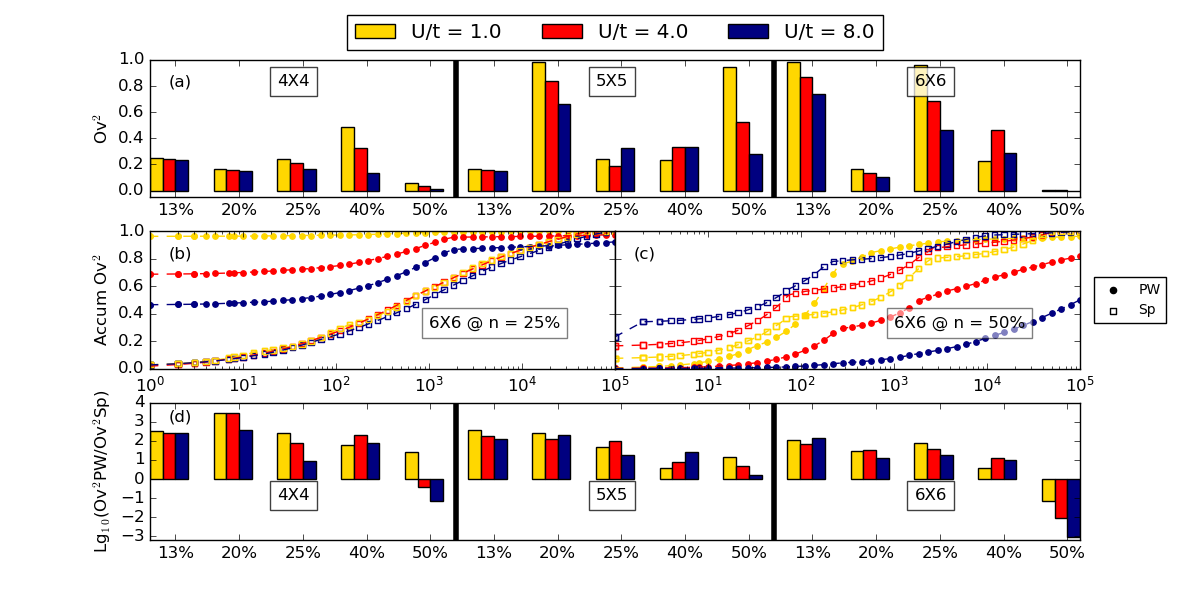}}
\end{center}
\caption{Results for the Hubbard model using ASCI with up to $10^7$ determinants for the plane wave (PW) results and up to $10^5$ determinants for the spatial basis (Sp) results. Here, particle fillings are giving as percentages, e.g. half-filling is given as $50\%$ and one-eighth filling as $12.5\% \approx 13\%$. \textbf{(a)} shows the squared overlap between the most important determinant and the full ASCI solution for 2D square lattice Hubbard models of different size (4x4, 5x5 and 6x6), ratio $U/t$ (from 1 to 8) and electron filling (from 13\% to 50 \%) using the PW basis. \textbf{(b)} and \textbf{(c)} show the squared overlap between the first $10^5$ determinants and the full ASCI solution for the 6x6 system at quarter and half filling respectively. The overlaps for the plane wave basis are shown in filled circles, the ones for the Sp basis in hollow squares. \textbf{(d)} shows the logarithm of the ratio of the squared overlap between the most important determinant and the full ASCI solution for the same 2D square lattice Hubbard model as in \textbf{(a)}.  \label{jar6}}
\label{fig:HubRes}
\end{figure*}

In Fig. \ref{fig:HubRes} we present an extensive set of results for the two-dimensional periodic square Hubbard model. 
In addition to varying the model parameters, we used here two sets of single-electron orbitals, a spatial representation and a Fourier representation. As the Hubbard Hamiltonian does not couple total momentum sectors, we chose to explore the zero-momentum space.

For small $U/t$ and filling, the single-determinant states in the plane wave basis have large overlap, Fig. \ref{fig:HubRes}(a), while single-determinant states in the spatial basis perform poorly, Fig. \ref{fig:HubRes}(d). \label{jar3} 
Increasing the filling at constant $U/t$ diminishes the overlap. For high $U/t$ at half-filling, where the exact ground state is an anti-ferromagnetic Mott insulator, the single-determinant state in the spatial basis has larger overlap than the single-determinant plane wave state in the 4x4 and 6x6 lattices. This phenomenon is not present in the 5x5 lattice due to frustration of the anti-ferromagnetic order.
To illustrate the efficiency of multi-reference states, we present in Fig. \ref{fig:HubRes}(b-c) the ground state overlap of multi-reference states for the 6x6 Hubbard model at quarter and half filling, for both plane wave and spatial bases. In general, about 10 determinants suffice for overlaps of about $5\%$.

\begin{figure}[htb!]
\begin{center}
\scalebox{1}{\includegraphics[width=\columnwidth]{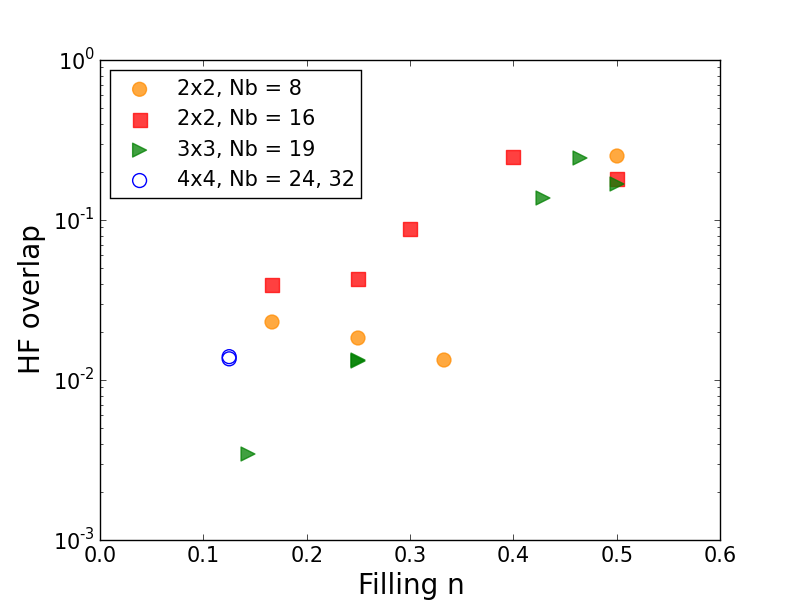}}
\end{center}
\caption{Averaged squared overlap between the most important single determinant and the full ASCI solution for Hubbard-Anderson models as arising in different cluster DMFT calculations on the square 2D Hubbard model for different cluster sizes and electron fillings. Here, the fillings are given as numbers, e.g. half-filling is shown as $0.5$, and one-eighth filling as $0.125$. The different DMFT calculations are: 2x2 cluster with 8 and 16 baths, 3x3 cluster with 19 baths, 4x4 cluster with 24 baths and 4x4 cluster with 32 baths. The two 4x4 calculations present identical overlaps.}.
\label{fig:Embed}
\end{figure}

In Fig. \ref{fig:Embed}, we explore the cluster DMFT \cite{Kotliar2001,Georges1996} solution for a 2D square lattice Hubbard model with $U/t = 8$ for various fillings, cluster sizes, and numbers of baths using the recently developed ASCI-DMFT algorithm \cite{Mejuto2017}. During the self-consistent iterations in the DMFT algorithm, multiple Hubbard-Anderson models are solved, and it is for this step that we consider performing state preparation and phase estimation.  
All calculations are performed in a basis of spatial orbitals.

Ground state overlaps decrease significantly with decreasing electron filling, Fig. \ref{fig:Embed}. This trend transcends cluster size and bath number, as seen by the fact that the two 4x4 points coincide, suggesting that this represents a physical property of the Hubbard model. Indeed, one can explain the overlap decrease between half and quarter filling by the disruption of the perfect anti-ferromagnetic order at half filling ($U/t = 8$ is relatively close to the Mott insulating phase \cite{auerbach:book}). The further decrease below quarter filling can then be ascribed to the inadequacy of a single determinant as a mean-field solution in the non-interacting limit. Thus, it appears that the use of a multi-determinant initial state will be particularly crucial for hybrid quantum-classical embedding algorithms. \label{ryan11}  

\section*{Conclusions} 

We have demonstrated by extensive tests on molecular systems, transition metal systems, Hubbard Models, the interacting electron gas, and embedding Hamiltonians that many classically intractable electronic structure problems will be amenable to efficient state preparation for phase estimation using single- and multi-determinant states. In conjunction with the quantum algorithm for multi-determinant state preparation, we have therefore addressed the two questions posed in the introduction and demonstrated that state preparation need not be a limiting factor in phase estimation even for large or strongly-correlated systems.

\section*{Acknowledgements}

This work was supported by the U.S. Department of Energy, Office of Science, Office of Advanced Scientific Computing Research, Quantum Algorithm Teams Program.
We used the Extreme Science and Engineering Discovery
Environment (XSEDE), which is supported by the National Science Foundation Grant No. OCI-1053575 and
resources of the Oak Ridge Leadership Computing Facility (OLCF) at the Oak Ridge National Laboratory,
which is supported by the Office of Science of the U.S.
Department of Energy under Contract No.  DE-AC0500OR22725. DH was funded by a Berkeley Fellowship. CMZ thanks the Fundaci\'on Bancaria La Caixa for a \emph{Obra Social ``La Caixa''} graduate fellowship. DL and MHG acknowledge support from the Director, Office of Science, Office of Basic Energy Sciences, of the U.S. Department of Energy under Contract No. DE-AC02-05CH11231. J.E. was supported by the Department of Defense (DoD) through the National Defense Science \& Engineering Graduate Fellowship (NDSEG) Program.



\begin{thebibliography}{78}%
\makeatletter
\providecommand \@ifxundefined [1]{%
 \@ifx{#1\undefined}
}%
\providecommand \@ifnum [1]{%
 \ifnum #1\expandafter \@firstoftwo
 \else \expandafter \@secondoftwo
 \fi
}%
\providecommand \@ifx [1]{%
 \ifx #1\expandafter \@firstoftwo
 \else \expandafter \@secondoftwo
 \fi
}%
\providecommand \natexlab [1]{#1}%
\providecommand \enquote  [1]{``#1''}%
\providecommand \bibnamefont  [1]{#1}%
\providecommand \bibfnamefont [1]{#1}%
\providecommand \citenamefont [1]{#1}%
\providecommand \href@noop [0]{\@secondoftwo}%
\providecommand \href [0]{\begingroup \@sanitize@url \@href}%
\providecommand \@href[1]{\@@startlink{#1}\@@href}%
\providecommand \@@href[1]{\endgroup#1\@@endlink}%
\providecommand \@sanitize@url [0]{\catcode `\\12\catcode `\$12\catcode
  `\&12\catcode `\#12\catcode `\^12\catcode `\_12\catcode `\%12\relax}%
\providecommand \@@startlink[1]{}%
\providecommand \@@endlink[0]{}%
\providecommand \url  [0]{\begingroup\@sanitize@url \@url }%
\providecommand \@url [1]{\endgroup\@href {#1}{\urlprefix }}%
\providecommand \urlprefix  [0]{URL }%
\providecommand \Eprint [0]{\href }%
\providecommand \doibase [0]{http://dx.doi.org/}%
\providecommand \selectlanguage [0]{\@gobble}%
\providecommand \bibinfo  [0]{\@secondoftwo}%
\providecommand \bibfield  [0]{\@secondoftwo}%
\providecommand \translation [1]{[#1]}%
\providecommand \BibitemOpen [0]{}%
\providecommand \bibitemStop [0]{}%
\providecommand \bibitemNoStop [0]{.\EOS\space}%
\providecommand \EOS [0]{\spacefactor3000\relax}%
\providecommand \BibitemShut  [1]{\csname bibitem#1\endcsname}%
\let\auto@bib@innerbib\@empty
\bibitem [{\citenamefont {Abrams}\ and\ \citenamefont
  {Lloyd}(1997)}]{Abrams1997}%
  \BibitemOpen
  \bibfield  {author} {\bibinfo {author} {\bibfnamefont {D.~S.}\ \bibnamefont
  {Abrams}}\ and\ \bibinfo {author} {\bibfnamefont {S.}~\bibnamefont {Lloyd}},\
  }\href {\doibase 10.1103/PhysRevLett.79.2586} {\bibfield  {journal} {\bibinfo
   {journal} {Phys. Rev. Lett.}\ }\textbf {\bibinfo {volume} {79}},\ \bibinfo
  {pages} {2586} (\bibinfo {year} {1997})}\BibitemShut {NoStop}%
\bibitem [{\citenamefont {Ortiz}\ \emph {et~al.}(2001)\citenamefont {Ortiz},
  \citenamefont {Gubernatis}, \citenamefont {Knill},\ and\ \citenamefont
  {Laflamme}}]{Ortiz2001}%
  \BibitemOpen
  \bibfield  {author} {\bibinfo {author} {\bibfnamefont {G.}~\bibnamefont
  {Ortiz}}, \bibinfo {author} {\bibfnamefont {J.}~\bibnamefont {Gubernatis}},
  \bibinfo {author} {\bibfnamefont {E.}~\bibnamefont {Knill}}, \ and\ \bibinfo
  {author} {\bibfnamefont {R.}~\bibnamefont {Laflamme}},\ }\href {\doibase
  10.1103/PhysRevA.64.022319} {\bibfield  {journal} {\bibinfo  {journal} {Phys.
  Rev. A}\ }\textbf {\bibinfo {volume} {64}},\ \bibinfo {pages} {22319}
  (\bibinfo {year} {2001})}\BibitemShut {NoStop}%
\bibitem [{\citenamefont {Aspuru-Guzik}\ \emph {et~al.}(2005)\citenamefont
  {Aspuru-Guzik}, \citenamefont {Dutoi}, \citenamefont {Love},\ and\
  \citenamefont {Head-Gordon}}]{guzik2005}%
  \BibitemOpen
  \bibfield  {author} {\bibinfo {author} {\bibfnamefont {A.}~\bibnamefont
  {Aspuru-Guzik}}, \bibinfo {author} {\bibfnamefont {A.~D.}\ \bibnamefont
  {Dutoi}}, \bibinfo {author} {\bibfnamefont {P.~J.}\ \bibnamefont {Love}}, \
  and\ \bibinfo {author} {\bibfnamefont {M.}~\bibnamefont {Head-Gordon}},\
  }\href {http://science.sciencemag.org/content/309/5741/1704.abstract}
  {\bibfield  {journal} {\bibinfo  {journal} {Science}\ }\textbf {\bibinfo
  {volume} {309}},\ \bibinfo {pages} {1704} (\bibinfo {year}
  {2005})}\BibitemShut {NoStop}%
\bibitem [{\citenamefont {McClean}\ \emph {et~al.}(2016)\citenamefont
  {McClean}, \citenamefont {Romero}, \citenamefont {Babbush},\ and\
  \citenamefont {Aspuru-Guzik}}]{mcclean2016}%
  \BibitemOpen
  \bibfield  {author} {\bibinfo {author} {\bibfnamefont {J.~R.}\ \bibnamefont
  {McClean}}, \bibinfo {author} {\bibfnamefont {J.}~\bibnamefont {Romero}},
  \bibinfo {author} {\bibfnamefont {R.}~\bibnamefont {Babbush}}, \ and\
  \bibinfo {author} {\bibfnamefont {A.}~\bibnamefont {Aspuru-Guzik}},\ }\href
  {http://stacks.iop.org/1367-2630/18/i=2/a=023023} {\bibfield  {journal}
  {\bibinfo  {journal} {New J. Phys.}\ }\textbf {\bibinfo {volume} {18}},\
  \bibinfo {pages} {023023} (\bibinfo {year} {2016})}\BibitemShut {NoStop}%
\bibitem [{\citenamefont {Babbush}\ \emph
  {et~al.}(2018{\natexlab{a}})\citenamefont {Babbush}, \citenamefont {Wiebe},
  \citenamefont {McClean}, \citenamefont {McClain}, \citenamefont {Neven},\
  and\ \citenamefont {Chan}}]{babbush2018}%
  \BibitemOpen
  \bibfield  {author} {\bibinfo {author} {\bibfnamefont {R.}~\bibnamefont
  {Babbush}}, \bibinfo {author} {\bibfnamefont {N.}~\bibnamefont {Wiebe}},
  \bibinfo {author} {\bibfnamefont {J.}~\bibnamefont {McClean}}, \bibinfo
  {author} {\bibfnamefont {J.}~\bibnamefont {McClain}}, \bibinfo {author}
  {\bibfnamefont {H.}~\bibnamefont {Neven}}, \ and\ \bibinfo {author}
  {\bibfnamefont {G.~K.-L.}\ \bibnamefont {Chan}},\ }\href {\doibase
  10.1103/PhysRevX.8.011044} {\bibfield  {journal} {\bibinfo  {journal} {Phys.
  Rev. X}\ }\textbf {\bibinfo {volume} {8}},\ \bibinfo {pages} {011044}
  (\bibinfo {year} {2018}{\natexlab{a}})}\BibitemShut {NoStop}%
\bibitem [{\citenamefont {Peruzzo}\ \emph {et~al.}(2014)\citenamefont
  {Peruzzo}, \citenamefont {McClean}, \citenamefont {Shadbolt}, \citenamefont
  {Yung}, \citenamefont {Zhou}, \citenamefont {Love}, \citenamefont
  {Aspuru-Guzik},\ and\ \citenamefont {O'Brien}}]{peruzzo2014}%
  \BibitemOpen
  \bibfield  {author} {\bibinfo {author} {\bibfnamefont {A.}~\bibnamefont
  {Peruzzo}}, \bibinfo {author} {\bibfnamefont {J.}~\bibnamefont {McClean}},
  \bibinfo {author} {\bibfnamefont {P.}~\bibnamefont {Shadbolt}}, \bibinfo
  {author} {\bibfnamefont {M.-H.}\ \bibnamefont {Yung}}, \bibinfo {author}
  {\bibfnamefont {X.-Q.}\ \bibnamefont {Zhou}}, \bibinfo {author}
  {\bibfnamefont {P.~J.}\ \bibnamefont {Love}}, \bibinfo {author}
  {\bibfnamefont {A.}~\bibnamefont {Aspuru-Guzik}}, \ and\ \bibinfo {author}
  {\bibfnamefont {J.~L.}\ \bibnamefont {O'Brien}},\ }\href
  {http://dx.doi.org/10.1038/ncomms5213} {\bibfield  {journal} {\bibinfo
  {journal} {Nat. Commun.}\ }\textbf {\bibinfo {volume} {5}},\ \bibinfo {pages}
  {4213 EP } (\bibinfo {year} {2014})}\BibitemShut {NoStop}%
\bibitem [{\citenamefont {Kivlichan}\ \emph {et~al.}(2018)\citenamefont
  {Kivlichan}, \citenamefont {McClean}, \citenamefont {Wiebe}, \citenamefont
  {Gidney}, \citenamefont {Aspuru-Guzik}, \citenamefont {Chan},\ and\
  \citenamefont {Babbush}}]{kivlichan2018quantum}%
  \BibitemOpen
  \bibfield  {author} {\bibinfo {author} {\bibfnamefont {I.~D.}\ \bibnamefont
  {Kivlichan}}, \bibinfo {author} {\bibfnamefont {J.}~\bibnamefont {McClean}},
  \bibinfo {author} {\bibfnamefont {N.}~\bibnamefont {Wiebe}}, \bibinfo
  {author} {\bibfnamefont {C.}~\bibnamefont {Gidney}}, \bibinfo {author}
  {\bibfnamefont {A.}~\bibnamefont {Aspuru-Guzik}}, \bibinfo {author}
  {\bibfnamefont {G.~K.-L.}\ \bibnamefont {Chan}}, \ and\ \bibinfo {author}
  {\bibfnamefont {R.}~\bibnamefont {Babbush}},\ }\href@noop {} {\bibfield
  {journal} {\bibinfo  {journal} {Phys. Rev. Lett.}\ }\textbf {\bibinfo
  {volume} {120}},\ \bibinfo {pages} {110501} (\bibinfo {year}
  {2018})}\BibitemShut {NoStop}%
\bibitem [{\citenamefont {{Kitaev}}(1995)}]{Kitaev1995}%
  \BibitemOpen
  \bibfield  {author} {\bibinfo {author} {\bibfnamefont {A.~Y.}\ \bibnamefont
  {{Kitaev}}},\ }\href@noop {} {\bibfield  {journal} {\bibinfo  {journal}
  {arXiv:quant-ph/9511026. Preprint, posted November 20, 1995.}\ } (\bibinfo
  {year} {1995})},\ \Eprint {http://arxiv.org/abs/quant-ph/9511026}
  {quant-ph/9511026} \BibitemShut {NoStop}%
\bibitem [{\citenamefont {Abrams}\ and\ \citenamefont
  {Lloyd}(1999)}]{Abrams1999}%
  \BibitemOpen
  \bibfield  {author} {\bibinfo {author} {\bibfnamefont {D.~S.}\ \bibnamefont
  {Abrams}}\ and\ \bibinfo {author} {\bibfnamefont {S.}~\bibnamefont {Lloyd}},\
  }\href {\doibase 10.1103/PhysRevLett.83.5162} {\bibfield  {journal} {\bibinfo
   {journal} {Phys. Rev. Lett.}\ }\textbf {\bibinfo {volume} {83}},\ \bibinfo
  {pages} {5162} (\bibinfo {year} {1999})}\BibitemShut {NoStop}%
\bibitem [{\citenamefont {Cody~Jones}\ \emph {et~al.}(2012)\citenamefont
  {Cody~Jones}, \citenamefont {Whitfield}, \citenamefont {McMahon},
  \citenamefont {Yung}, \citenamefont {Meter}, \citenamefont {Aspuru-Guzik},\
  and\ \citenamefont {Yamamoto}}]{Jones2012}%
  \BibitemOpen
  \bibfield  {author} {\bibinfo {author} {\bibfnamefont {N.}~\bibnamefont
  {Cody~Jones}}, \bibinfo {author} {\bibfnamefont {J.~D.}\ \bibnamefont
  {Whitfield}}, \bibinfo {author} {\bibfnamefont {P.~L.}\ \bibnamefont
  {McMahon}}, \bibinfo {author} {\bibfnamefont {M.-H.}\ \bibnamefont {Yung}},
  \bibinfo {author} {\bibfnamefont {R.~V.}\ \bibnamefont {Meter}}, \bibinfo
  {author} {\bibfnamefont {A.}~\bibnamefont {Aspuru-Guzik}}, \ and\ \bibinfo
  {author} {\bibfnamefont {Y.}~\bibnamefont {Yamamoto}},\ }\href {\doibase
  10.1088/1367-2630/14/11/115023} {\bibfield  {journal} {\bibinfo  {journal}
  {New J. Phys.}\ }\textbf {\bibinfo {volume} {14}},\ \bibinfo {pages} {115023}
  (\bibinfo {year} {2012})}\BibitemShut {NoStop}%
\bibitem [{\citenamefont {Wecker}\ \emph {et~al.}(2014)\citenamefont {Wecker},
  \citenamefont {Bauer}, \citenamefont {Clark}, \citenamefont {Hastings},\ and\
  \citenamefont {Troyer}}]{wecker2014}%
  \BibitemOpen
  \bibfield  {author} {\bibinfo {author} {\bibfnamefont {D.}~\bibnamefont
  {Wecker}}, \bibinfo {author} {\bibfnamefont {B.}~\bibnamefont {Bauer}},
  \bibinfo {author} {\bibfnamefont {B.~K.}\ \bibnamefont {Clark}}, \bibinfo
  {author} {\bibfnamefont {M.~B.}\ \bibnamefont {Hastings}}, \ and\ \bibinfo
  {author} {\bibfnamefont {M.}~\bibnamefont {Troyer}},\ }\href {\doibase
  10.1103/PhysRevA.90.022305} {\bibfield  {journal} {\bibinfo  {journal} {Phys.
  Rev. A}\ }\textbf {\bibinfo {volume} {90}},\ \bibinfo {pages} {022305}
  (\bibinfo {year} {2014})}\BibitemShut {NoStop}%
\bibitem [{\citenamefont {Reiher}\ \emph {et~al.}(2017)\citenamefont {Reiher},
  \citenamefont {Wiebe}, \citenamefont {Svore}, \citenamefont {Wecker},\ and\
  \citenamefont {Troyer}}]{Reiher2017}%
  \BibitemOpen
  \bibfield  {author} {\bibinfo {author} {\bibfnamefont {M.}~\bibnamefont
  {Reiher}}, \bibinfo {author} {\bibfnamefont {N.}~\bibnamefont {Wiebe}},
  \bibinfo {author} {\bibfnamefont {K.~M.}\ \bibnamefont {Svore}}, \bibinfo
  {author} {\bibfnamefont {D.}~\bibnamefont {Wecker}}, \ and\ \bibinfo {author}
  {\bibfnamefont {M.}~\bibnamefont {Troyer}},\ }\href@noop {} {\bibfield
  {journal} {\bibinfo  {journal} {Proc. Natl. Acad. Sci.}\ }\textbf {\bibinfo
  {volume} {114}},\ \bibinfo {pages} {7555} (\bibinfo {year}
  {2017})}\BibitemShut {NoStop}%
\bibitem [{\citenamefont {Babbush}\ \emph
  {et~al.}(2018{\natexlab{b}})\citenamefont {Babbush}, \citenamefont {Gidney},
  \citenamefont {Berry}, \citenamefont {Wiebe}, \citenamefont {McClean},
  \citenamefont {Paler}, \citenamefont {Fowler},\ and\ \citenamefont
  {Neven}}]{BabbushSpectra}%
  \BibitemOpen
  \bibfield  {author} {\bibinfo {author} {\bibfnamefont {R.}~\bibnamefont
  {Babbush}}, \bibinfo {author} {\bibfnamefont {C.}~\bibnamefont {Gidney}},
  \bibinfo {author} {\bibfnamefont {D.}~\bibnamefont {Berry}}, \bibinfo
  {author} {\bibfnamefont {N.}~\bibnamefont {Wiebe}}, \bibinfo {author}
  {\bibfnamefont {J.}~\bibnamefont {McClean}}, \bibinfo {author} {\bibfnamefont
  {A.}~\bibnamefont {Paler}}, \bibinfo {author} {\bibfnamefont
  {A.}~\bibnamefont {Fowler}}, \ and\ \bibinfo {author} {\bibfnamefont
  {H.}~\bibnamefont {Neven}},\ }\href {https://arxiv.org/abs/1805.03662}
  {\bibfield  {journal} {\bibinfo  {journal} {arXiv:1805.03662v1. Preprint,
  posted May 9, 2018}\ } (\bibinfo {year} {2018}{\natexlab{b}})}\BibitemShut
  {NoStop}%
\bibitem [{\citenamefont {Lanyon}\ \emph {et~al.}(2010)\citenamefont {Lanyon},
  \citenamefont {Whitfield}, \citenamefont {Gillett}, \citenamefont {Goggin},
  \citenamefont {Almeida}, \citenamefont {Kassal}, \citenamefont {Biamonte},
  \citenamefont {Mohseni}, \citenamefont {Powell}, \citenamefont {Barbieri}
  \emph {et~al.}}]{lanyon2010towards}%
  \BibitemOpen
  \bibfield  {author} {\bibinfo {author} {\bibfnamefont {B.~P.}\ \bibnamefont
  {Lanyon}}, \bibinfo {author} {\bibfnamefont {J.~D.}\ \bibnamefont
  {Whitfield}}, \bibinfo {author} {\bibfnamefont {G.~G.}\ \bibnamefont
  {Gillett}}, \bibinfo {author} {\bibfnamefont {M.~E.}\ \bibnamefont {Goggin}},
  \bibinfo {author} {\bibfnamefont {M.~P.}\ \bibnamefont {Almeida}}, \bibinfo
  {author} {\bibfnamefont {I.}~\bibnamefont {Kassal}}, \bibinfo {author}
  {\bibfnamefont {J.~D.}\ \bibnamefont {Biamonte}}, \bibinfo {author}
  {\bibfnamefont {M.}~\bibnamefont {Mohseni}}, \bibinfo {author} {\bibfnamefont
  {B.~J.}\ \bibnamefont {Powell}}, \bibinfo {author} {\bibfnamefont
  {M.}~\bibnamefont {Barbieri}},  \emph {et~al.},\ }\href@noop {} {\bibfield
  {journal} {\bibinfo  {journal} {Nature chemistry}\ }\textbf {\bibinfo
  {volume} {2}},\ \bibinfo {pages} {106} (\bibinfo {year} {2010})}\BibitemShut
  {NoStop}%
\bibitem [{\citenamefont {Wang}\ \emph {et~al.}(2015)\citenamefont {Wang},
  \citenamefont {Dolde}, \citenamefont {Biamonte}, \citenamefont {Babbush},
  \citenamefont {Bergholm}, \citenamefont {Yang}, \citenamefont {Jakobi},
  \citenamefont {Neumann}, \citenamefont {Aspuru-Guzik}, \citenamefont
  {Whitfield},\ and\ \citenamefont {Wrachtrup}}]{Wang2014}%
  \BibitemOpen
  \bibfield  {author} {\bibinfo {author} {\bibfnamefont {Y.}~\bibnamefont
  {Wang}}, \bibinfo {author} {\bibfnamefont {F.}~\bibnamefont {Dolde}},
  \bibinfo {author} {\bibfnamefont {J.}~\bibnamefont {Biamonte}}, \bibinfo
  {author} {\bibfnamefont {R.}~\bibnamefont {Babbush}}, \bibinfo {author}
  {\bibfnamefont {V.}~\bibnamefont {Bergholm}}, \bibinfo {author}
  {\bibfnamefont {S.}~\bibnamefont {Yang}}, \bibinfo {author} {\bibfnamefont
  {I.}~\bibnamefont {Jakobi}}, \bibinfo {author} {\bibfnamefont
  {P.}~\bibnamefont {Neumann}}, \bibinfo {author} {\bibfnamefont
  {A.}~\bibnamefont {Aspuru-Guzik}}, \bibinfo {author} {\bibfnamefont {J.~D.}\
  \bibnamefont {Whitfield}}, \ and\ \bibinfo {author} {\bibfnamefont
  {J.}~\bibnamefont {Wrachtrup}},\ }\href {\doibase 10.1021/acsnano.5b01651}
  {\bibfield  {journal} {\bibinfo  {journal} {ACS Nano}\ }\textbf {\bibinfo
  {volume} {9}},\ \bibinfo {pages} {7769} (\bibinfo {year} {2015})}\BibitemShut
  {NoStop}%
\bibitem [{\citenamefont {O'Malley}\ \emph {et~al.}(2016)\citenamefont
  {O'Malley}, \citenamefont {Babbush}, \citenamefont {Kivlichan}, \citenamefont
  {Romero}, \citenamefont {McClean}, \citenamefont {Barends}, \citenamefont
  {Kelly}, \citenamefont {Roushan}, \citenamefont {Tranter}, \citenamefont
  {Ding}, \citenamefont {Campbell}, \citenamefont {Chen}, \citenamefont {Chen},
  \citenamefont {Chiaro}, \citenamefont {Dunsworth}, \citenamefont {Fowler},
  \citenamefont {Jeffrey}, \citenamefont {Megrant}, \citenamefont {Mutus},
  \citenamefont {Neill}, \citenamefont {Quintana}, \citenamefont {Sank},
  \citenamefont {Vainsencher}, \citenamefont {Wenner}, \citenamefont {White},
  \citenamefont {Coveney}, \citenamefont {Love}, \citenamefont {Neven},
  \citenamefont {Aspuru-Guzik},\ and\ \citenamefont {Martinis}}]{OMalley2016}%
  \BibitemOpen
  \bibfield  {author} {\bibinfo {author} {\bibfnamefont {P.~J.~J.}\
  \bibnamefont {O'Malley}}, \bibinfo {author} {\bibfnamefont {R.}~\bibnamefont
  {Babbush}}, \bibinfo {author} {\bibfnamefont {I.~D.}\ \bibnamefont
  {Kivlichan}}, \bibinfo {author} {\bibfnamefont {J.}~\bibnamefont {Romero}},
  \bibinfo {author} {\bibfnamefont {J.~R.}\ \bibnamefont {McClean}}, \bibinfo
  {author} {\bibfnamefont {R.}~\bibnamefont {Barends}}, \bibinfo {author}
  {\bibfnamefont {J.}~\bibnamefont {Kelly}}, \bibinfo {author} {\bibfnamefont
  {P.}~\bibnamefont {Roushan}}, \bibinfo {author} {\bibfnamefont
  {A.}~\bibnamefont {Tranter}}, \bibinfo {author} {\bibfnamefont
  {N.}~\bibnamefont {Ding}}, \bibinfo {author} {\bibfnamefont {B.}~\bibnamefont
  {Campbell}}, \bibinfo {author} {\bibfnamefont {Y.}~\bibnamefont {Chen}},
  \bibinfo {author} {\bibfnamefont {Z.}~\bibnamefont {Chen}}, \bibinfo {author}
  {\bibfnamefont {B.}~\bibnamefont {Chiaro}}, \bibinfo {author} {\bibfnamefont
  {A.}~\bibnamefont {Dunsworth}}, \bibinfo {author} {\bibfnamefont {A.~G.}\
  \bibnamefont {Fowler}}, \bibinfo {author} {\bibfnamefont {E.}~\bibnamefont
  {Jeffrey}}, \bibinfo {author} {\bibfnamefont {A.}~\bibnamefont {Megrant}},
  \bibinfo {author} {\bibfnamefont {J.~Y.}\ \bibnamefont {Mutus}}, \bibinfo
  {author} {\bibfnamefont {C.}~\bibnamefont {Neill}}, \bibinfo {author}
  {\bibfnamefont {C.}~\bibnamefont {Quintana}}, \bibinfo {author}
  {\bibfnamefont {D.}~\bibnamefont {Sank}}, \bibinfo {author} {\bibfnamefont
  {A.}~\bibnamefont {Vainsencher}}, \bibinfo {author} {\bibfnamefont
  {J.}~\bibnamefont {Wenner}}, \bibinfo {author} {\bibfnamefont {T.~C.}\
  \bibnamefont {White}}, \bibinfo {author} {\bibfnamefont {P.~V.}\ \bibnamefont
  {Coveney}}, \bibinfo {author} {\bibfnamefont {P.~J.}\ \bibnamefont {Love}},
  \bibinfo {author} {\bibfnamefont {H.}~\bibnamefont {Neven}}, \bibinfo
  {author} {\bibfnamefont {A.}~\bibnamefont {Aspuru-Guzik}}, \ and\ \bibinfo
  {author} {\bibfnamefont {J.~M.}\ \bibnamefont {Martinis}},\ }\href {\doibase
  http://dx.doi.org/10.1103/PhysRevX.6.031007} {\bibfield  {journal} {\bibinfo
  {journal} {Phys. Rev. X}\ }\textbf {\bibinfo {volume} {6}},\ \bibinfo {pages}
  {31007} (\bibinfo {year} {2016})}\BibitemShut {NoStop}%
\bibitem [{\citenamefont {Whitfield}\ \emph {et~al.}(2011)\citenamefont
  {Whitfield}, \citenamefont {Biamonte},\ and\ \citenamefont
  {Aspuru-Guzik}}]{Whitfield2010}%
  \BibitemOpen
  \bibfield  {author} {\bibinfo {author} {\bibfnamefont {J.~D.}\ \bibnamefont
  {Whitfield}}, \bibinfo {author} {\bibfnamefont {J.}~\bibnamefont {Biamonte}},
  \ and\ \bibinfo {author} {\bibfnamefont {A.}~\bibnamefont {Aspuru-Guzik}},\
  }\href {\doibase 10.1080/00268976.2011.552441} {\bibfield  {journal}
  {\bibinfo  {journal} {Mol. Phys.}\ }\textbf {\bibinfo {volume} {109}},\
  \bibinfo {pages} {735} (\bibinfo {year} {2011})}\BibitemShut {NoStop}%
\bibitem [{\citenamefont {Kohn}(1999)}]{Kohn1999nobel}%
  \BibitemOpen
  \bibfield  {author} {\bibinfo {author} {\bibfnamefont {W.}~\bibnamefont
  {Kohn}},\ }\href@noop {} {\bibfield  {journal} {\bibinfo  {journal} {Rev.
  Mod. Phys.}\ }\textbf {\bibinfo {volume} {71}},\ \bibinfo {pages} {1253}
  (\bibinfo {year} {1999})}\BibitemShut {NoStop}%
\bibitem [{\citenamefont {Bartlett}\ \emph {et~al.}(1989)\citenamefont
  {Bartlett}, \citenamefont {Kucharski},\ and\ \citenamefont
  {Noga}}]{Bartlett1989}%
  \BibitemOpen
  \bibfield  {author} {\bibinfo {author} {\bibfnamefont {R.~J.}\ \bibnamefont
  {Bartlett}}, \bibinfo {author} {\bibfnamefont {S.~A.}\ \bibnamefont
  {Kucharski}}, \ and\ \bibinfo {author} {\bibfnamefont {J.}~\bibnamefont
  {Noga}},\ }\href {\doibase 10.1016/S0009-2614(89)87372-5} {\bibfield
  {journal} {\bibinfo  {journal} {Chem. Phys. Lett.}\ }\textbf {\bibinfo
  {volume} {155}},\ \bibinfo {pages} {133} (\bibinfo {year}
  {1989})}\BibitemShut {NoStop}%
\bibitem [{\citenamefont {Romero}\ \emph {et~al.}(2017)\citenamefont {Romero},
  \citenamefont {Babbush}, \citenamefont {McClean}, \citenamefont {Hempel},
  \citenamefont {Love},\ and\ \citenamefont {Aspuru-Guzik}}]{Romero2017}%
  \BibitemOpen
  \bibfield  {author} {\bibinfo {author} {\bibfnamefont {J.}~\bibnamefont
  {Romero}}, \bibinfo {author} {\bibfnamefont {R.}~\bibnamefont {Babbush}},
  \bibinfo {author} {\bibfnamefont {J.}~\bibnamefont {McClean}}, \bibinfo
  {author} {\bibfnamefont {C.}~\bibnamefont {Hempel}}, \bibinfo {author}
  {\bibfnamefont {P.}~\bibnamefont {Love}}, \ and\ \bibinfo {author}
  {\bibfnamefont {A.}~\bibnamefont {Aspuru-Guzik}},\ }\href
  {http://arxiv.org/abs/1701.02691} {\bibfield  {journal} {\bibinfo  {journal}
  {arXiv:1701.02691}\ } (\bibinfo {year} {2017})}\BibitemShut {NoStop}%
\bibitem [{\citenamefont {Wu}\ \emph {et~al.}(2002)\citenamefont {Wu},
  \citenamefont {Byrd},\ and\ \citenamefont {Lidar}}]{Wu2002}%
  \BibitemOpen
  \bibfield  {author} {\bibinfo {author} {\bibfnamefont {L.-A.}\ \bibnamefont
  {Wu}}, \bibinfo {author} {\bibfnamefont {M.~S.}\ \bibnamefont {Byrd}}, \ and\
  \bibinfo {author} {\bibfnamefont {D.~A.}\ \bibnamefont {Lidar}},\ }\href
  {\doibase 10.1103/PhysRevLett.89.057904} {\bibfield  {journal} {\bibinfo
  {journal} {Phys. Rev. Lett.}\ }\textbf {\bibinfo {volume} {89}},\ \bibinfo
  {pages} {057904} (\bibinfo {year} {2002})}\BibitemShut {NoStop}%
\bibitem [{\citenamefont {Babbush}\ \emph {et~al.}(2014)\citenamefont
  {Babbush}, \citenamefont {Love},\ and\ \citenamefont
  {Aspuru-Guzik}}]{BabbushAQChem}%
  \BibitemOpen
  \bibfield  {author} {\bibinfo {author} {\bibfnamefont {R.}~\bibnamefont
  {Babbush}}, \bibinfo {author} {\bibfnamefont {P.~J.}\ \bibnamefont {Love}}, \
  and\ \bibinfo {author} {\bibfnamefont {A.}~\bibnamefont {Aspuru-Guzik}},\
  }\href {\doibase 10.1038/srep06603} {\bibfield  {journal} {\bibinfo
  {journal} {Sci. Rep.-UK}\ }\textbf {\bibinfo {volume} {4}},\ \bibinfo {pages}
  {1} (\bibinfo {year} {2014})}\BibitemShut {NoStop}%
\bibitem [{\citenamefont {Ward}\ \emph {et~al.}(2008)\citenamefont {Ward},
  \citenamefont {Kassal},\ and\ \citenamefont {Aspuru-Guzik}}]{Ward2009}%
  \BibitemOpen
  \bibfield  {author} {\bibinfo {author} {\bibfnamefont {N.~J.}\ \bibnamefont
  {Ward}}, \bibinfo {author} {\bibfnamefont {I.}~\bibnamefont {Kassal}}, \ and\
  \bibinfo {author} {\bibfnamefont {A.}~\bibnamefont {Aspuru-Guzik}},\ }\href
  {\doibase http://dx.doi.org/10.1063/1.3115177} {\bibfield  {journal}
  {\bibinfo  {journal} {J. Of Chem. Phys.}\ }\textbf {\bibinfo {volume}
  {130}},\ \bibinfo {pages} {194105} (\bibinfo {year} {2008})}\BibitemShut
  {NoStop}%
\bibitem [{\citenamefont {McClean}\ \emph {et~al.}(2014)\citenamefont
  {McClean}, \citenamefont {Babbush}, \citenamefont {Love},\ and\ \citenamefont
  {Aspuru-Guzik}}]{McClean2014}%
  \BibitemOpen
  \bibfield  {author} {\bibinfo {author} {\bibfnamefont {J.~R.}\ \bibnamefont
  {McClean}}, \bibinfo {author} {\bibfnamefont {R.}~\bibnamefont {Babbush}},
  \bibinfo {author} {\bibfnamefont {P.~J.}\ \bibnamefont {Love}}, \ and\
  \bibinfo {author} {\bibfnamefont {A.}~\bibnamefont {Aspuru-Guzik}},\ }\href
  {\doibase 10.1021/jz501649m} {\bibfield  {journal} {\bibinfo  {journal} {J.
  Phys. Chem. Lett.}\ }\textbf {\bibinfo {volume} {5}},\ \bibinfo {pages}
  {4368} (\bibinfo {year} {2014})}\BibitemShut {NoStop}%
\bibitem [{\citenamefont {Tubman}\ \emph {et~al.}(2016)\citenamefont {Tubman},
  \citenamefont {Lee}, \citenamefont {Takeshita}, \citenamefont {Head-Gordon},\
  and\ \citenamefont {Whaley}}]{tubman2016-1}%
  \BibitemOpen
  \bibfield  {author} {\bibinfo {author} {\bibfnamefont {N.~M.}\ \bibnamefont
  {Tubman}}, \bibinfo {author} {\bibfnamefont {J.}~\bibnamefont {Lee}},
  \bibinfo {author} {\bibfnamefont {T.~Y.}\ \bibnamefont {Takeshita}}, \bibinfo
  {author} {\bibfnamefont {M.}~\bibnamefont {Head-Gordon}}, \ and\ \bibinfo
  {author} {\bibfnamefont {K.~B.}\ \bibnamefont {Whaley}},\ }\href {\doibase
  http://dx.doi.org/10.1063/1.4955109} {\bibfield  {journal} {\bibinfo
  {journal} {J. Chem. Phys.}\ }\textbf {\bibinfo {volume} {145}},\ \bibinfo
  {eid} {044112} (\bibinfo {year} {2016}),\
  http://dx.doi.org/10.1063/1.4955109}\BibitemShut {NoStop}%
\bibitem [{\citenamefont {L\"auchli}\ \emph {et~al.}(2011)\citenamefont
  {L\"auchli}, \citenamefont {Sudan},\ and\ \citenamefont
  {S\o{}rensen}}]{lauchli2011}%
  \BibitemOpen
  \bibfield  {author} {\bibinfo {author} {\bibfnamefont {A.~M.}\ \bibnamefont
  {L\"auchli}}, \bibinfo {author} {\bibfnamefont {J.}~\bibnamefont {Sudan}}, \
  and\ \bibinfo {author} {\bibfnamefont {E.~S.}\ \bibnamefont {S\o{}rensen}},\
  }\href {\doibase 10.1103/PhysRevB.83.212401} {\bibfield  {journal} {\bibinfo
  {journal} {Phys. Rev. B}\ }\textbf {\bibinfo {volume} {83}},\ \bibinfo
  {pages} {212401} (\bibinfo {year} {2011})}\BibitemShut {NoStop}%
\bibitem [{\citenamefont {Vogiatzis}\ \emph {et~al.}(2017)\citenamefont
  {Vogiatzis}, \citenamefont {Ma}, \citenamefont {Olsen}, \citenamefont
  {Gagliardi},\ and\ \citenamefont {de~Jong}}]{vogiatzis2017}%
  \BibitemOpen
  \bibfield  {author} {\bibinfo {author} {\bibfnamefont {K.~D.}\ \bibnamefont
  {Vogiatzis}}, \bibinfo {author} {\bibfnamefont {D.}~\bibnamefont {Ma}},
  \bibinfo {author} {\bibfnamefont {J.}~\bibnamefont {Olsen}}, \bibinfo
  {author} {\bibfnamefont {L.}~\bibnamefont {Gagliardi}}, \ and\ \bibinfo
  {author} {\bibfnamefont {W.~A.}\ \bibnamefont {de~Jong}},\ }\href {\doibase
  10.1063/1.4989858} {\bibfield  {journal} {\bibinfo  {journal} {J. Chem.
  Phys.}\ }\textbf {\bibinfo {volume} {147}},\ \bibinfo {pages} {184111}
  (\bibinfo {year} {2017})},\ \Eprint
  {http://arxiv.org/abs/https://doi.org/10.1063/1.4989858}
  {https://doi.org/10.1063/1.4989858} \BibitemShut {NoStop}%
\bibitem [{\citenamefont {Szabo}\ and\ \citenamefont
  {Ostlund}(1982)}]{szabo:book}%
  \BibitemOpen
  \bibfield  {author} {\bibinfo {author} {\bibfnamefont {A.}~\bibnamefont
  {Szabo}}\ and\ \bibinfo {author} {\bibfnamefont {N.}~\bibnamefont
  {Ostlund}},\ }\href@noop {} {\emph {\bibinfo {title} {{M}odern {Q}uantum
  {C}hemistry}}}\ (\bibinfo  {publisher} {Dover},\ \bibinfo {year}
  {1982})\BibitemShut {NoStop}%
\bibitem [{\citenamefont {Gan}\ \emph {et~al.}(2006)\citenamefont {Gan},
  \citenamefont {Grant}, \citenamefont {Harrison},\ and\ \citenamefont
  {Dixon}}]{gan2006}%
  \BibitemOpen
  \bibfield  {author} {\bibinfo {author} {\bibfnamefont {Z.}~\bibnamefont
  {Gan}}, \bibinfo {author} {\bibfnamefont {D.~J.}\ \bibnamefont {Grant}},
  \bibinfo {author} {\bibfnamefont {R.~J.}\ \bibnamefont {Harrison}}, \ and\
  \bibinfo {author} {\bibfnamefont {D.~A.}\ \bibnamefont {Dixon}},\ }\href
  {http://scitation.aip.org/content/aip/journal/jcp/125/12/10.1063/1.2335446}
  {\bibfield  {journal} {\bibinfo  {journal} {J. Chem. Phys.}\ }\textbf
  {\bibinfo {volume} {125}},\ \bibinfo {eid} {124311} (\bibinfo {year}
  {2006})}\BibitemShut {NoStop}%
\bibitem [{\citenamefont {Gan}\ and\ \citenamefont {Harrison}(2005)}]{gan2005}%
  \BibitemOpen
  \bibfield  {author} {\bibinfo {author} {\bibfnamefont {Z.}~\bibnamefont
  {Gan}}\ and\ \bibinfo {author} {\bibfnamefont {R.}~\bibnamefont {Harrison}},\
  }in\ \href {\doibase 10.1109/SC.2005.17} {\emph {\bibinfo {booktitle}
  {Supercomputing, 2005. Proceedings of the ACM/IEEE SC 2005 Conference}}}\
  (\bibinfo {year} {2005})\ pp.\ \bibinfo {pages} {22--22}\BibitemShut
  {NoStop}%
\bibitem [{\citenamefont {Sherrill}\ and\ \citenamefont
  {III.}(1999)}]{sherrill1999}%
  \BibitemOpen
  \bibfield  {author} {\bibinfo {author} {\bibfnamefont {C.~D.}\ \bibnamefont
  {Sherrill}}\ and\ \bibinfo {author} {\bibfnamefont {H.~F.~S.}\ \bibnamefont
  {III.}}\ }(\bibinfo  {publisher} {Academic Press},\ \bibinfo {year} {1999})\
  pp.\ \bibinfo {pages} {143 -- 269}\BibitemShut {NoStop}%
\bibitem [{\citenamefont {Abrams}\ and\ \citenamefont
  {Sherrill}(2004)}]{abrams2004}%
  \BibitemOpen
  \bibfield  {author} {\bibinfo {author} {\bibfnamefont {M.~L.}\ \bibnamefont
  {Abrams}}\ and\ \bibinfo {author} {\bibfnamefont {C.~D.}\ \bibnamefont
  {Sherrill}},\ }\href {\doibase http://dx.doi.org/10.1063/1.1804498}
  {\bibfield  {journal} {\bibinfo  {journal} {J. Chem. Phys.}\ }\textbf
  {\bibinfo {volume} {121}},\ \bibinfo {pages} {9211} (\bibinfo {year}
  {2004})}\BibitemShut {NoStop}%
\bibitem [{\citenamefont {Szalay}\ \emph {et~al.}(2012)\citenamefont {Szalay},
  \citenamefont {Muller}, \citenamefont {Gidofalvi}, \citenamefont {Lischka},\
  and\ \citenamefont {Shepard}}]{szalay2012}%
  \BibitemOpen
  \bibfield  {author} {\bibinfo {author} {\bibfnamefont {P.~G.}\ \bibnamefont
  {Szalay}}, \bibinfo {author} {\bibfnamefont {T.}~\bibnamefont {Muller}},
  \bibinfo {author} {\bibfnamefont {G.}~\bibnamefont {Gidofalvi}}, \bibinfo
  {author} {\bibfnamefont {H.}~\bibnamefont {Lischka}}, \ and\ \bibinfo
  {author} {\bibfnamefont {R.}~\bibnamefont {Shepard}},\ }\href {\doibase
  10.1021/cr200137a} {\bibfield  {journal} {\bibinfo  {journal} {Chem. Rev.}\
  }\textbf {\bibinfo {volume} {112}},\ \bibinfo {pages} {108} (\bibinfo {year}
  {2012})}\BibitemShut {NoStop}%
\bibitem [{\citenamefont {Bender}\ and\ \citenamefont
  {Davidson}(1969)}]{bender1969}%
  \BibitemOpen
  \bibfield  {author} {\bibinfo {author} {\bibfnamefont {C.~F.}\ \bibnamefont
  {Bender}}\ and\ \bibinfo {author} {\bibfnamefont {E.~R.}\ \bibnamefont
  {Davidson}},\ }\href {\doibase 10.1103/PhysRev.183.23} {\bibfield  {journal}
  {\bibinfo  {journal} {Phys. Rev.}\ }\textbf {\bibinfo {volume} {183}},\
  \bibinfo {pages} {23} (\bibinfo {year} {1969})}\BibitemShut {NoStop}%
\bibitem [{\citenamefont {Buenker}\ \emph {et~al.}(1978)\citenamefont
  {Buenker}, \citenamefont {Peyerimhoff},\ and\ \citenamefont
  {Butscher}}]{buenker1978}%
  \BibitemOpen
  \bibfield  {author} {\bibinfo {author} {\bibfnamefont {R.~J.}\ \bibnamefont
  {Buenker}}, \bibinfo {author} {\bibfnamefont {S.~D.}\ \bibnamefont
  {Peyerimhoff}}, \ and\ \bibinfo {author} {\bibfnamefont {W.}~\bibnamefont
  {Butscher}},\ }\href {\doibase 10.1080/00268977800100581} {\bibfield
  {journal} {\bibinfo  {journal} {Mol. Phys.}\ }\textbf {\bibinfo {volume}
  {35}},\ \bibinfo {pages} {771} (\bibinfo {year} {1978})}\BibitemShut
  {NoStop}%
\bibitem [{\citenamefont {Roth}(2009)}]{roth2009}%
  \BibitemOpen
  \bibfield  {author} {\bibinfo {author} {\bibfnamefont {R.}~\bibnamefont
  {Roth}},\ }\href {\doibase 10.1103/PhysRevC.79.064324} {\bibfield  {journal}
  {\bibinfo  {journal} {Phys. Rev. C}\ }\textbf {\bibinfo {volume} {79}},\
  \bibinfo {pages} {064324} (\bibinfo {year} {2009})}\BibitemShut {NoStop}%
\bibitem [{\citenamefont {Illas}\ \emph {et~al.}(1991)\citenamefont {Illas},
  \citenamefont {Rubio}, \citenamefont {Ricart},\ and\ \citenamefont
  {Bagus}}]{bagus1991}%
  \BibitemOpen
  \bibfield  {author} {\bibinfo {author} {\bibfnamefont {F.}~\bibnamefont
  {Illas}}, \bibinfo {author} {\bibfnamefont {J.}~\bibnamefont {Rubio}},
  \bibinfo {author} {\bibfnamefont {J.~M.}\ \bibnamefont {Ricart}}, \ and\
  \bibinfo {author} {\bibfnamefont {P.~S.}\ \bibnamefont {Bagus}},\ }\href
  {\doibase http://dx.doi.org/10.1063/1.461037} {\bibfield  {journal} {\bibinfo
   {journal} {J. Chem. Phys.}\ }\textbf {\bibinfo {volume} {95}},\ \bibinfo
  {pages} {1877} (\bibinfo {year} {1991})}\BibitemShut {NoStop}%
\bibitem [{\citenamefont {Huron}\ \emph {et~al.}(1973)\citenamefont {Huron},
  \citenamefont {Malrieu},\ and\ \citenamefont {Rancurel}}]{huron1973}%
  \BibitemOpen
  \bibfield  {author} {\bibinfo {author} {\bibfnamefont {B.}~\bibnamefont
  {Huron}}, \bibinfo {author} {\bibfnamefont {J.~P.}\ \bibnamefont {Malrieu}},
  \ and\ \bibinfo {author} {\bibfnamefont {P.}~\bibnamefont {Rancurel}},\
  }\href
  {http://scitation.aip.org/content/aip/journal/jcp/58/12/10.1063/1.1679199}
  {\bibfield  {journal} {\bibinfo  {journal} {J. Chem. Phys.}\ }\textbf
  {\bibinfo {volume} {58}},\ \bibinfo {pages} {5745} (\bibinfo {year}
  {1973})}\BibitemShut {NoStop}%
\bibitem [{\citenamefont {Wecker}\ \emph {et~al.}(2015)\citenamefont {Wecker},
  \citenamefont {Hastings}, \citenamefont {Wiebe}, \citenamefont {Clark},
  \citenamefont {Nayak},\ and\ \citenamefont {Troyer}}]{wecker2015solving}%
  \BibitemOpen
  \bibfield  {author} {\bibinfo {author} {\bibfnamefont {D.}~\bibnamefont
  {Wecker}}, \bibinfo {author} {\bibfnamefont {M.~B.}\ \bibnamefont
  {Hastings}}, \bibinfo {author} {\bibfnamefont {N.}~\bibnamefont {Wiebe}},
  \bibinfo {author} {\bibfnamefont {B.~K.}\ \bibnamefont {Clark}}, \bibinfo
  {author} {\bibfnamefont {C.}~\bibnamefont {Nayak}}, \ and\ \bibinfo {author}
  {\bibfnamefont {M.}~\bibnamefont {Troyer}},\ }\href@noop {} {\bibfield
  {journal} {\bibinfo  {journal} {Phys. Rev. A}\ }\textbf {\bibinfo {volume}
  {92}},\ \bibinfo {pages} {062318} (\bibinfo {year} {2015})}\BibitemShut
  {NoStop}%
\bibitem [{\citenamefont {Shende}\ \emph {et~al.}(2006)\citenamefont {Shende},
  \citenamefont {Bullock},\ and\ \citenamefont
  {Markov}}]{Shende_synthesis_2006}%
  \BibitemOpen
  \bibfield  {author} {\bibinfo {author} {\bibfnamefont {V.~V.}\ \bibnamefont
  {Shende}}, \bibinfo {author} {\bibfnamefont {S.~S.}\ \bibnamefont {Bullock}},
  \ and\ \bibinfo {author} {\bibfnamefont {I.~L.}\ \bibnamefont {Markov}},\
  }\href {\doibase 10.1109/TCAD.2005.855930} {\bibfield  {journal} {\bibinfo
  {journal} {IEEE Transactions on Computer-Aided Design of Integrated Circuits
  and Systems}\ }\textbf {\bibinfo {volume} {25}},\ \bibinfo {pages} {1000}
  (\bibinfo {year} {2006})}\BibitemShut {NoStop}%
\bibitem [{\citenamefont {Childs}\ \emph {et~al.}(2017)\citenamefont {Childs},
  \citenamefont {Maslov}, \citenamefont {Nam}, \citenamefont {Ross},\ and\
  \citenamefont {Su}}]{Childs_Maslov_Nam_Ross_Su_2017}%
  \BibitemOpen
  \bibfield  {author} {\bibinfo {author} {\bibfnamefont {A.~M.}\ \bibnamefont
  {Childs}}, \bibinfo {author} {\bibfnamefont {D.}~\bibnamefont {Maslov}},
  \bibinfo {author} {\bibfnamefont {Y.}~\bibnamefont {Nam}}, \bibinfo {author}
  {\bibfnamefont {N.~J.}\ \bibnamefont {Ross}}, \ and\ \bibinfo {author}
  {\bibfnamefont {Y.}~\bibnamefont {Su}},\ }\href
  {http://arxiv.org/abs/1711.10980} {\bibfield  {journal} {\bibinfo  {journal}
  {arXiv:1711.10980}\ } (\bibinfo {year} {2017})}\BibitemShut {NoStop}%
\bibitem [{\citenamefont {Pople}\ \emph {et~al.}(1989)\citenamefont {Pople},
  \citenamefont {Gordon}, \citenamefont {Fox}, \citenamefont {Raghavachari},\
  and\ \citenamefont {Curtiss}}]{pople1989}%
  \BibitemOpen
  \bibfield  {author} {\bibinfo {author} {\bibfnamefont {J.~A.}\ \bibnamefont
  {Pople}}, \bibinfo {author} {\bibfnamefont {M.~H.}\ \bibnamefont {Gordon}},
  \bibinfo {author} {\bibfnamefont {D.~J.}\ \bibnamefont {Fox}}, \bibinfo
  {author} {\bibfnamefont {K.}~\bibnamefont {Raghavachari}}, \ and\ \bibinfo
  {author} {\bibfnamefont {L.~A.}\ \bibnamefont {Curtiss}},\ }\href {\doibase
  10.1063/1.456415} {\bibfield  {journal} {\bibinfo  {journal} {J. Chem.
  Phys.}\ }\textbf {\bibinfo {volume} {90}},\ \bibinfo {pages} {5622} (\bibinfo
  {year} {1989})},\ \Eprint
  {http://arxiv.org/abs/https://doi.org/10.1063/1.456415}
  {https://doi.org/10.1063/1.456415} \BibitemShut {NoStop}%
\bibitem [{\citenamefont {Tubman}\ \emph
  {et~al.}(2018{\natexlab{a}})\citenamefont {Tubman}, \citenamefont {Freeman},
  \citenamefont {Levine}, \citenamefont {Hait}, \citenamefont {Head-Gordon},\
  and\ \citenamefont {Whaley}}]{tubman2018-1}%
  \BibitemOpen
  \bibfield  {author} {\bibinfo {author} {\bibfnamefont {N.~M.}\ \bibnamefont
  {Tubman}}, \bibinfo {author} {\bibfnamefont {C.~D.}\ \bibnamefont {Freeman}},
  \bibinfo {author} {\bibfnamefont {D.~S.}\ \bibnamefont {Levine}}, \bibinfo
  {author} {\bibfnamefont {D.}~\bibnamefont {Hait}}, \bibinfo {author}
  {\bibfnamefont {M.}~\bibnamefont {Head-Gordon}}, \ and\ \bibinfo {author}
  {\bibfnamefont {K.~B.}\ \bibnamefont {Whaley}},\ }\href@noop {} {\bibfield
  {journal} {\bibinfo  {journal} {arXiv:1807.00821v1. Preprint, posted July 2,
  2018}\ } (\bibinfo {year} {2018}{\natexlab{a}})}\BibitemShut {NoStop}%
\bibitem [{\citenamefont {Tubman}\ \emph
  {et~al.}(2018{\natexlab{b}})\citenamefont {Tubman}, \citenamefont {Levine},
  \citenamefont {Hait}, \citenamefont {Head-Gordon},\ and\ \citenamefont
  {Whaley}}]{tubman2018-2}%
  \BibitemOpen
  \bibfield  {author} {\bibinfo {author} {\bibfnamefont {N.~M.}\ \bibnamefont
  {Tubman}}, \bibinfo {author} {\bibfnamefont {D.~S.}\ \bibnamefont {Levine}},
  \bibinfo {author} {\bibfnamefont {D.}~\bibnamefont {Hait}}, \bibinfo {author}
  {\bibfnamefont {M.}~\bibnamefont {Head-Gordon}}, \ and\ \bibinfo {author}
  {\bibfnamefont {K.~B.}\ \bibnamefont {Whaley}},\ }\href@noop {} {\bibfield
  {journal} {\bibinfo  {journal} {arXiv:1808.02049v1. Preprint, posted August
  6, 2018.}\ } (\bibinfo {year} {2018}{\natexlab{b}})}\BibitemShut {NoStop}%
\bibitem [{\citenamefont {Feller}\ \emph {et~al.}(2008)\citenamefont {Feller},
  \citenamefont {Peterson},\ and\ \citenamefont {Dixon}}]{feller1}%
  \BibitemOpen
  \bibfield  {author} {\bibinfo {author} {\bibfnamefont {D.}~\bibnamefont
  {Feller}}, \bibinfo {author} {\bibfnamefont {K.~A.}\ \bibnamefont
  {Peterson}}, \ and\ \bibinfo {author} {\bibfnamefont {D.~A.}\ \bibnamefont
  {Dixon}},\ }\href {\doibase 10.1063/1.3008061} {\bibfield  {journal}
  {\bibinfo  {journal} {J. Chem. Phys.}\ }\textbf {\bibinfo {volume} {129}},\
  \bibinfo {pages} {204105} (\bibinfo {year} {2008})},\ \Eprint
  {http://arxiv.org/abs/https://doi.org/10.1063/1.3008061}
  {https://doi.org/10.1063/1.3008061} \BibitemShut {NoStop}%
\bibitem [{\citenamefont {Mardirossian}\ and\ \citenamefont
  {Head-Gordon}(2017)}]{mardirossian2017thirty}%
  \BibitemOpen
  \bibfield  {author} {\bibinfo {author} {\bibfnamefont {N.}~\bibnamefont
  {Mardirossian}}\ and\ \bibinfo {author} {\bibfnamefont {M.}~\bibnamefont
  {Head-Gordon}},\ }\href@noop {} {\bibfield  {journal} {\bibinfo  {journal}
  {Mol. Phys.}\ }\textbf {\bibinfo {volume} {115}},\ \bibinfo {pages} {2315}
  (\bibinfo {year} {2017})}\BibitemShut {NoStop}%
\bibitem [{\citenamefont {Hait}\ and\ \citenamefont
  {Head-Gordon}(2018)}]{hait2018accurate}%
  \BibitemOpen
  \bibfield  {author} {\bibinfo {author} {\bibfnamefont {D.}~\bibnamefont
  {Hait}}\ and\ \bibinfo {author} {\bibfnamefont {M.}~\bibnamefont
  {Head-Gordon}},\ }\href@noop {} {\bibfield  {journal} {\bibinfo  {journal}
  {J. Chem. Theory Comput.}\ }\textbf {\bibinfo {volume} {14}},\ \bibinfo
  {pages} {1969} (\bibinfo {year} {2018})}\BibitemShut {NoStop}%
\bibitem [{\citenamefont {Gwaltney}\ \emph {et~al.}(2002)\citenamefont
  {Gwaltney}, \citenamefont {Byrd}, \citenamefont {Van~Voorhis},\ and\
  \citenamefont {Head-Gordon}}]{gwaltney2002perturbative}%
  \BibitemOpen
  \bibfield  {author} {\bibinfo {author} {\bibfnamefont {S.~R.}\ \bibnamefont
  {Gwaltney}}, \bibinfo {author} {\bibfnamefont {E.~F.}\ \bibnamefont {Byrd}},
  \bibinfo {author} {\bibfnamefont {T.}~\bibnamefont {Van~Voorhis}}, \ and\
  \bibinfo {author} {\bibfnamefont {M.}~\bibnamefont {Head-Gordon}},\
  }\href@noop {} {\bibfield  {journal} {\bibinfo  {journal} {Chem. Phys.
  Lett.}\ }\textbf {\bibinfo {volume} {353}},\ \bibinfo {pages} {359} (\bibinfo
  {year} {2002})}\BibitemShut {NoStop}%
\bibitem [{\citenamefont {Van~Voorhis}\ and\ \citenamefont
  {Head-Gordon}(2000)}]{van2000quadratic}%
  \BibitemOpen
  \bibfield  {author} {\bibinfo {author} {\bibfnamefont {T.}~\bibnamefont
  {Van~Voorhis}}\ and\ \bibinfo {author} {\bibfnamefont {M.}~\bibnamefont
  {Head-Gordon}},\ }\href@noop {} {\bibfield  {journal} {\bibinfo  {journal}
  {Chem. Phys. Lett.}\ }\textbf {\bibinfo {volume} {330}},\ \bibinfo {pages}
  {585} (\bibinfo {year} {2000})}\BibitemShut {NoStop}%
\bibitem [{\citenamefont {Sharma}\ \emph {et~al.}(2017)\citenamefont {Sharma},
  \citenamefont {Holmes}, \citenamefont {Jeanmairet}, \citenamefont {Alavi},\
  and\ \citenamefont {Umrigar}}]{sharma2017}%
  \BibitemOpen
  \bibfield  {author} {\bibinfo {author} {\bibfnamefont {S.}~\bibnamefont
  {Sharma}}, \bibinfo {author} {\bibfnamefont {A.~A.}\ \bibnamefont {Holmes}},
  \bibinfo {author} {\bibfnamefont {G.}~\bibnamefont {Jeanmairet}}, \bibinfo
  {author} {\bibfnamefont {A.}~\bibnamefont {Alavi}}, \ and\ \bibinfo {author}
  {\bibfnamefont {C.~J.}\ \bibnamefont {Umrigar}},\ }\href {\doibase
  10.1021/acs.jctc.6b01028} {\bibfield  {journal} {\bibinfo  {journal} {J.
  Chem. Theory Comput.}\ }\textbf {\bibinfo {volume} {13}},\ \bibinfo {pages}
  {1595} (\bibinfo {year} {2017})},\ \bibinfo {note} {pMID: 28263594},\ \Eprint
  {http://arxiv.org/abs/https://doi.org/10.1021/acs.jctc.6b01028}
  {https://doi.org/10.1021/acs.jctc.6b01028} \BibitemShut {NoStop}%
\bibitem [{\citenamefont {Gull}\ \emph {et~al.}(2010)\citenamefont {Gull},
  \citenamefont {Ferrero}, \citenamefont {Parcollet}, \citenamefont {Georges},\
  and\ \citenamefont {Millis}}]{Gull2010}%
  \BibitemOpen
  \bibfield  {author} {\bibinfo {author} {\bibfnamefont {E.}~\bibnamefont
  {Gull}}, \bibinfo {author} {\bibfnamefont {M.}~\bibnamefont {Ferrero}},
  \bibinfo {author} {\bibfnamefont {O.}~\bibnamefont {Parcollet}}, \bibinfo
  {author} {\bibfnamefont {A.}~\bibnamefont {Georges}}, \ and\ \bibinfo
  {author} {\bibfnamefont {A.~J.}\ \bibnamefont {Millis}},\ }\href@noop {}
  {\bibfield  {journal} {\bibinfo  {journal} {Phys. Rev. B}\ }\textbf {\bibinfo
  {volume} {82}},\ \bibinfo {pages} {155101} (\bibinfo {year}
  {2010})}\BibitemShut {NoStop}%
\bibitem [{\citenamefont {Sordi}\ \emph {et~al.}(2012)\citenamefont {Sordi},
  \citenamefont {S\'emon}, \citenamefont {Haule},\ and\ \citenamefont
  {Tremblay}}]{Sordi2012}%
  \BibitemOpen
  \bibfield  {author} {\bibinfo {author} {\bibfnamefont {G.}~\bibnamefont
  {Sordi}}, \bibinfo {author} {\bibfnamefont {P.}~\bibnamefont {S\'emon}},
  \bibinfo {author} {\bibfnamefont {K.}~\bibnamefont {Haule}}, \ and\ \bibinfo
  {author} {\bibfnamefont {A.-M.~S.}\ \bibnamefont {Tremblay}},\ }\href@noop {}
  {\bibfield  {journal} {\bibinfo  {journal} {Phys. Rev. Lett.}\ }\textbf
  {\bibinfo {volume} {108}},\ \bibinfo {pages} {216401} (\bibinfo {year}
  {2012})}\BibitemShut {NoStop}%
\bibitem [{\citenamefont {Zheng}\ \emph {et~al.}(2017)\citenamefont {Zheng},
  \citenamefont {Chung}, \citenamefont {Corboz}, \citenamefont {Ehlers},
  \citenamefont {Qin}, \citenamefont {Noack}, \citenamefont {Shi},
  \citenamefont {White}, \citenamefont {Zhang},\ and\ \citenamefont
  {Chan}}]{Zheng2017}%
  \BibitemOpen
  \bibfield  {author} {\bibinfo {author} {\bibfnamefont {B.-X.}\ \bibnamefont
  {Zheng}}, \bibinfo {author} {\bibfnamefont {C.-M.}\ \bibnamefont {Chung}},
  \bibinfo {author} {\bibfnamefont {P.}~\bibnamefont {Corboz}}, \bibinfo
  {author} {\bibfnamefont {G.}~\bibnamefont {Ehlers}}, \bibinfo {author}
  {\bibfnamefont {M.-P.}\ \bibnamefont {Qin}}, \bibinfo {author} {\bibfnamefont
  {R.~M.}\ \bibnamefont {Noack}}, \bibinfo {author} {\bibfnamefont
  {H.}~\bibnamefont {Shi}}, \bibinfo {author} {\bibfnamefont {S.~R.}\
  \bibnamefont {White}}, \bibinfo {author} {\bibfnamefont {S.}~\bibnamefont
  {Zhang}}, \ and\ \bibinfo {author} {\bibfnamefont {G.~K.-L.}\ \bibnamefont
  {Chan}},\ }\href@noop {} {\bibfield  {journal} {\bibinfo  {journal}
  {Science}\ }\textbf {\bibinfo {volume} {358}},\ \bibinfo {pages} {1155}
  (\bibinfo {year} {2017})}\BibitemShut {NoStop}%
\bibitem [{\citenamefont {Acharya}\ \emph {et~al.}(2018)\citenamefont
  {Acharya}, \citenamefont {Weber}, \citenamefont {Plekhanov}, \citenamefont
  {Pashov}, \citenamefont {Taraphder},\ and\ \citenamefont
  {Schilfgaarde}}]{Acharya2018}%
  \BibitemOpen
  \bibfield  {author} {\bibinfo {author} {\bibfnamefont {S.}~\bibnamefont
  {Acharya}}, \bibinfo {author} {\bibfnamefont {C.}~\bibnamefont {Weber}},
  \bibinfo {author} {\bibfnamefont {E.}~\bibnamefont {Plekhanov}}, \bibinfo
  {author} {\bibfnamefont {D.}~\bibnamefont {Pashov}}, \bibinfo {author}
  {\bibfnamefont {A.}~\bibnamefont {Taraphder}}, \ and\ \bibinfo {author}
  {\bibfnamefont {M.~V.}\ \bibnamefont {Schilfgaarde}},\ }\href@noop {}
  {\bibfield  {journal} {\bibinfo  {journal} {Phys. Rev. X}\ }\textbf {\bibinfo
  {volume} {8}},\ \bibinfo {pages} {021038} (\bibinfo {year}
  {2018})}\BibitemShut {NoStop}%
\bibitem [{\citenamefont {Hwang}\ \emph {et~al.}(0)\citenamefont {Hwang},
  \citenamefont {Kim}, \citenamefont {Ryu}, \citenamefont {Kim}, \citenamefont
  {Lee}, \citenamefont {Kim}, \citenamefont {Kang}, \citenamefont {Park},
  \citenamefont {Lanzara}, \citenamefont {Chung}, \citenamefont {Mo},
  \citenamefont {Denlinger}, \citenamefont {Min},\ and\ \citenamefont
  {Hwang}}]{Hwang2018}%
  \BibitemOpen
  \bibfield  {author} {\bibinfo {author} {\bibfnamefont {J.}~\bibnamefont
  {Hwang}}, \bibinfo {author} {\bibfnamefont {K.}~\bibnamefont {Kim}}, \bibinfo
  {author} {\bibfnamefont {H.}~\bibnamefont {Ryu}}, \bibinfo {author}
  {\bibfnamefont {J.}~\bibnamefont {Kim}}, \bibinfo {author} {\bibfnamefont
  {J.-E.}\ \bibnamefont {Lee}}, \bibinfo {author} {\bibfnamefont
  {S.}~\bibnamefont {Kim}}, \bibinfo {author} {\bibfnamefont {M.}~\bibnamefont
  {Kang}}, \bibinfo {author} {\bibfnamefont {B.-G.}\ \bibnamefont {Park}},
  \bibinfo {author} {\bibfnamefont {A.}~\bibnamefont {Lanzara}}, \bibinfo
  {author} {\bibfnamefont {J.}~\bibnamefont {Chung}}, \bibinfo {author}
  {\bibfnamefont {S.-K.}\ \bibnamefont {Mo}}, \bibinfo {author} {\bibfnamefont
  {J.}~\bibnamefont {Denlinger}}, \bibinfo {author} {\bibfnamefont {B.~I.}\
  \bibnamefont {Min}}, \ and\ \bibinfo {author} {\bibfnamefont
  {C.}~\bibnamefont {Hwang}},\ }\href {\doibase 10.1021/acs.nanolett.8b00784}
  {\bibfield  {journal} {\bibinfo  {journal} {Nano Letters}\ }\textbf {\bibinfo
  {volume} {0}},\ \bibinfo {pages} {null} (\bibinfo {year} {0})},\ \bibinfo
  {note} {pMID: 29761696},\ \Eprint
  {http://arxiv.org/abs/https://doi.org/10.1021/acs.nanolett.8b00784}
  {https://doi.org/10.1021/acs.nanolett.8b00784} \BibitemShut {NoStop}%
\bibitem [{\citenamefont {Bauer}\ \emph {et~al.}(2016)\citenamefont {Bauer},
  \citenamefont {Wecker}, \citenamefont {Millis}, \citenamefont {Hastings},\
  and\ \citenamefont {Troyer}}]{Bauer2016}%
  \BibitemOpen
  \bibfield  {author} {\bibinfo {author} {\bibfnamefont {B.}~\bibnamefont
  {Bauer}}, \bibinfo {author} {\bibfnamefont {D.}~\bibnamefont {Wecker}},
  \bibinfo {author} {\bibfnamefont {A.~J.}\ \bibnamefont {Millis}}, \bibinfo
  {author} {\bibfnamefont {M.~B.}\ \bibnamefont {Hastings}}, \ and\ \bibinfo
  {author} {\bibfnamefont {M.}~\bibnamefont {Troyer}},\ }\href {\doibase
  10.1103/PhysRevX.6.031045} {\bibfield  {journal} {\bibinfo  {journal} {Phys.
  Rev. X}\ }\textbf {\bibinfo {volume} {6}},\ \bibinfo {pages} {031045}
  (\bibinfo {year} {2016})}\BibitemShut {NoStop}%
\bibitem [{\citenamefont {Sakai}\ \emph {et~al.}(2009)\citenamefont {Sakai},
  \citenamefont {Motome},\ and\ \citenamefont {Imada}}]{Sakai2009}%
  \BibitemOpen
  \bibfield  {author} {\bibinfo {author} {\bibfnamefont {S.}~\bibnamefont
  {Sakai}}, \bibinfo {author} {\bibfnamefont {Y.}~\bibnamefont {Motome}}, \
  and\ \bibinfo {author} {\bibfnamefont {M.}~\bibnamefont {Imada}},\
  }\href@noop {} {\bibfield  {journal} {\bibinfo  {journal} {Phys. Rev. Lett.}\
  }\textbf {\bibinfo {volume} {102}},\ \bibinfo {pages} {056404} (\bibinfo
  {year} {2009})}\BibitemShut {NoStop}%
\bibitem [{\citenamefont {Ehlers}\ \emph {et~al.}(2017)\citenamefont {Ehlers},
  \citenamefont {White},\ and\ \citenamefont {Noack}}]{Ehlers2017}%
  \BibitemOpen
  \bibfield  {author} {\bibinfo {author} {\bibfnamefont {G.}~\bibnamefont
  {Ehlers}}, \bibinfo {author} {\bibfnamefont {S.~R.}\ \bibnamefont {White}}, \
  and\ \bibinfo {author} {\bibfnamefont {R.~M.}\ \bibnamefont {Noack}},\
  }\href@noop {} {\bibfield  {journal} {\bibinfo  {journal} {Phys. Rev. B}\
  }\textbf {\bibinfo {volume} {95}},\ \bibinfo {pages} {125125} (\bibinfo
  {year} {2017})}\BibitemShut {NoStop}%
\bibitem [{\citenamefont {Ceperley}\ and\ \citenamefont
  {Alder}(1980)}]{ceperley1980}%
  \BibitemOpen
  \bibfield  {author} {\bibinfo {author} {\bibfnamefont {D.~M.}\ \bibnamefont
  {Ceperley}}\ and\ \bibinfo {author} {\bibfnamefont {B.~J.}\ \bibnamefont
  {Alder}},\ }\href {\doibase 10.1103/PhysRevLett.45.566} {\bibfield  {journal}
  {\bibinfo  {journal} {Phys. Rev. Lett.}\ }\textbf {\bibinfo {volume} {45}},\
  \bibinfo {pages} {566} (\bibinfo {year} {1980})}\BibitemShut {NoStop}%
\bibitem [{\citenamefont {Vosko}\ \emph {et~al.}(1980)\citenamefont {Vosko},
  \citenamefont {Wilk},\ and\ \citenamefont {Nusair}}]{VWN}%
  \BibitemOpen
  \bibfield  {author} {\bibinfo {author} {\bibfnamefont {S.~H.}\ \bibnamefont
  {Vosko}}, \bibinfo {author} {\bibfnamefont {L.}~\bibnamefont {Wilk}}, \ and\
  \bibinfo {author} {\bibfnamefont {M.}~\bibnamefont {Nusair}},\ }\href@noop {}
  {\bibfield  {journal} {\bibinfo  {journal} {Can. J. Phys.}\ }\textbf
  {\bibinfo {volume} {58}},\ \bibinfo {pages} {1200} (\bibinfo {year}
  {1980})}\BibitemShut {NoStop}%
\bibitem [{\citenamefont {Perdew}\ and\ \citenamefont {Wang}(1992)}]{PW92}%
  \BibitemOpen
  \bibfield  {author} {\bibinfo {author} {\bibfnamefont {J.~P.}\ \bibnamefont
  {Perdew}}\ and\ \bibinfo {author} {\bibfnamefont {Y.}~\bibnamefont {Wang}},\
  }\href@noop {} {\bibfield  {journal} {\bibinfo  {journal} {Phys. Rev. B}\
  }\textbf {\bibinfo {volume} {45}},\ \bibinfo {pages} {13244} (\bibinfo {year}
  {1992})}\BibitemShut {NoStop}%
\bibitem [{\citenamefont {Shepherd}\ \emph {et~al.}(2012)\citenamefont
  {Shepherd}, \citenamefont {Booth},\ and\ \citenamefont
  {Alavi}}]{shepherd2012investigation}%
  \BibitemOpen
  \bibfield  {author} {\bibinfo {author} {\bibfnamefont {J.~J.}\ \bibnamefont
  {Shepherd}}, \bibinfo {author} {\bibfnamefont {G.~H.}\ \bibnamefont {Booth}},
  \ and\ \bibinfo {author} {\bibfnamefont {A.}~\bibnamefont {Alavi}},\
  }\href@noop {} {\bibfield  {journal} {\bibinfo  {journal} {J. Chem. Phys.}\
  }\textbf {\bibinfo {volume} {136}},\ \bibinfo {pages} {244101} (\bibinfo
  {year} {2012})}\BibitemShut {NoStop}%
\bibitem [{\citenamefont {Bernu}\ \emph {et~al.}(2011)\citenamefont {Bernu},
  \citenamefont {Delyon}, \citenamefont {Holzmann},\ and\ \citenamefont
  {Baguet}}]{bernu2011}%
  \BibitemOpen
  \bibfield  {author} {\bibinfo {author} {\bibfnamefont {B.}~\bibnamefont
  {Bernu}}, \bibinfo {author} {\bibfnamefont {F.}~\bibnamefont {Delyon}},
  \bibinfo {author} {\bibfnamefont {M.}~\bibnamefont {Holzmann}}, \ and\
  \bibinfo {author} {\bibfnamefont {L.}~\bibnamefont {Baguet}},\ }\href
  {\doibase 10.1103/PhysRevB.84.115115} {\bibfield  {journal} {\bibinfo
  {journal} {Phys. Rev. B}\ }\textbf {\bibinfo {volume} {84}},\ \bibinfo
  {pages} {115115} (\bibinfo {year} {2011})}\BibitemShut {NoStop}%
\bibitem [{\citenamefont {Brown}\ \emph {et~al.}(2013)\citenamefont {Brown},
  \citenamefont {Clark}, \citenamefont {DuBois},\ and\ \citenamefont
  {Ceperley}}]{brown2013}%
  \BibitemOpen
  \bibfield  {author} {\bibinfo {author} {\bibfnamefont {E.~W.}\ \bibnamefont
  {Brown}}, \bibinfo {author} {\bibfnamefont {B.~K.}\ \bibnamefont {Clark}},
  \bibinfo {author} {\bibfnamefont {J.~L.}\ \bibnamefont {DuBois}}, \ and\
  \bibinfo {author} {\bibfnamefont {D.~M.}\ \bibnamefont {Ceperley}},\ }\href
  {\doibase 10.1103/PhysRevLett.110.146405} {\bibfield  {journal} {\bibinfo
  {journal} {Phys. Rev. Lett.}\ }\textbf {\bibinfo {volume} {110}},\ \bibinfo
  {pages} {146405} (\bibinfo {year} {2013})}\BibitemShut {NoStop}%
\bibitem [{\citenamefont {Malone}\ \emph {et~al.}(2016)\citenamefont {Malone},
  \citenamefont {Blunt}, \citenamefont {Brown}, \citenamefont {Lee},
  \citenamefont {Spencer}, \citenamefont {Foulkes},\ and\ \citenamefont
  {Shepherd}}]{malone2016accurate}%
  \BibitemOpen
  \bibfield  {author} {\bibinfo {author} {\bibfnamefont {F.~D.}\ \bibnamefont
  {Malone}}, \bibinfo {author} {\bibfnamefont {N.}~\bibnamefont {Blunt}},
  \bibinfo {author} {\bibfnamefont {E.~W.}\ \bibnamefont {Brown}}, \bibinfo
  {author} {\bibfnamefont {D.}~\bibnamefont {Lee}}, \bibinfo {author}
  {\bibfnamefont {J.}~\bibnamefont {Spencer}}, \bibinfo {author} {\bibfnamefont
  {W.}~\bibnamefont {Foulkes}}, \ and\ \bibinfo {author} {\bibfnamefont
  {J.~J.}\ \bibnamefont {Shepherd}},\ }\href@noop {} {\bibfield  {journal}
  {\bibinfo  {journal} {Phys. Rev. Lett.}\ }\textbf {\bibinfo {volume} {117}},\
  \bibinfo {pages} {115701} (\bibinfo {year} {2016})}\BibitemShut {NoStop}%
\bibitem [{\citenamefont {Roos}\ \emph {et~al.}(1980)\citenamefont {Roos},
  \citenamefont {Taylor}, \citenamefont {Si} \emph
  {et~al.}}]{roos1980complete}%
  \BibitemOpen
  \bibfield  {author} {\bibinfo {author} {\bibfnamefont {B.~O.}\ \bibnamefont
  {Roos}}, \bibinfo {author} {\bibfnamefont {P.~R.}\ \bibnamefont {Taylor}},
  \bibinfo {author} {\bibfnamefont {P.~E.}\ \bibnamefont {Si}},  \emph
  {et~al.},\ }\href@noop {} {\bibfield  {journal} {\bibinfo  {journal}
  {Chemical Physics}\ }\textbf {\bibinfo {volume} {48}},\ \bibinfo {pages}
  {157} (\bibinfo {year} {1980})}\BibitemShut {NoStop}%
\bibitem [{\citenamefont {Levine}\ \emph {et~al.}(2018)\citenamefont {Levine}
  \emph {et~al.}}]{levine2018}%
  \BibitemOpen
  \bibfield  {author} {\bibinfo {author} {\bibfnamefont {D.}~\bibnamefont
  {Levine}} \emph {et~al.},\ }\href@noop {} {\bibfield  {journal} {\bibinfo
  {journal} {In preparation}\ } (\bibinfo {year} {2018})}\BibitemShut {NoStop}%
\bibitem [{\citenamefont {Kotliar}\ \emph {et~al.}(2001)\citenamefont
  {Kotliar}, \citenamefont {Savrasov}, \citenamefont {Palsson},\ and\
  \citenamefont {Biroli}}]{Kotliar2001}%
  \BibitemOpen
  \bibfield  {author} {\bibinfo {author} {\bibfnamefont {G.}~\bibnamefont
  {Kotliar}}, \bibinfo {author} {\bibfnamefont {S.~Y.}\ \bibnamefont
  {Savrasov}}, \bibinfo {author} {\bibfnamefont {G.}~\bibnamefont {Palsson}}, \
  and\ \bibinfo {author} {\bibfnamefont {G.}~\bibnamefont {Biroli}},\
  }\href@noop {} {\bibfield  {journal} {\bibinfo  {journal} {Phys. Rev. Lett.}\
  }\textbf {\bibinfo {volume} {87}},\ \bibinfo {pages} {186401} (\bibinfo
  {year} {2001})}\BibitemShut {NoStop}%
\bibitem [{\citenamefont {Georges}\ \emph
  {et~al.}(1996{\natexlab{a}})\citenamefont {Georges}, \citenamefont {Kotliar},
  \citenamefont {Krauth},\ and\ \citenamefont {Rozenberg}}]{Georges1996}%
  \BibitemOpen
  \bibfield  {author} {\bibinfo {author} {\bibfnamefont {A.}~\bibnamefont
  {Georges}}, \bibinfo {author} {\bibfnamefont {G.}~\bibnamefont {Kotliar}},
  \bibinfo {author} {\bibfnamefont {W.}~\bibnamefont {Krauth}}, \ and\ \bibinfo
  {author} {\bibfnamefont {M.~J.}\ \bibnamefont {Rozenberg}},\ }\href@noop {}
  {\bibfield  {journal} {\bibinfo  {journal} {Rev. Mod. Phys.}\ }\textbf
  {\bibinfo {volume} {68}},\ \bibinfo {pages} {13} (\bibinfo {year}
  {1996}{\natexlab{a}})}\BibitemShut {NoStop}%
\bibitem [{\citenamefont {Mejuto-Zaera}\ \emph {et~al.}(2017)\citenamefont
  {Mejuto-Zaera}, \citenamefont {Tubman},\ and\ \citenamefont
  {Whaley}}]{Mejuto2017}%
  \BibitemOpen
  \bibfield  {author} {\bibinfo {author} {\bibfnamefont {C.}~\bibnamefont
  {Mejuto-Zaera}}, \bibinfo {author} {\bibfnamefont {N.~M.}\ \bibnamefont
  {Tubman}}, \ and\ \bibinfo {author} {\bibfnamefont {K.~B.}\ \bibnamefont
  {Whaley}},\ }\href@noop {} {\bibfield  {journal} {\bibinfo  {journal}
  {arXiv:1711.04771v2. Preprint, posted November 22, 2017}\ } (\bibinfo {year}
  {2017})},\ \Eprint {http://arxiv.org/abs/1711.04771} {arXiv:1711.04771
  [cond-mat]} \BibitemShut {NoStop}%
\bibitem [{\citenamefont {Auerbach}(1994)}]{auerbach:book}%
  \BibitemOpen
  \bibfield  {author} {\bibinfo {author} {\bibfnamefont {A.}~\bibnamefont
  {Auerbach}},\ }\href@noop {} {\emph {\bibinfo {title} {{I}nteracting
  {E}lectrons and {Q}uantum {M}agnetism}}}\ (\bibinfo  {publisher} {Springer},\
  \bibinfo {year} {1994})\BibitemShut {NoStop}%
\bibitem [{\citenamefont {Shao}\ \emph {et~al.}(2015)\citenamefont {Shao} \emph
  {et~al.}}]{shao_advances_2015}%
  \BibitemOpen
  \bibfield  {author} {\bibinfo {author} {\bibfnamefont {Y.}~\bibnamefont
  {Shao}} \emph {et~al.},\ }\href {\doibase 10.1080/00268976.2014.952696}
  {\bibfield  {journal} {\bibinfo  {journal} {Mol. Phys.}\ }\textbf {\bibinfo
  {volume} {113}},\ \bibinfo {pages} {184} (\bibinfo {year}
  {2015})}\BibitemShut {NoStop}%
\bibitem [{\citenamefont {Dunning}(1989)}]{Dunning1989}%
  \BibitemOpen
  \bibfield  {author} {\bibinfo {author} {\bibfnamefont {T.~H.}\ \bibnamefont
  {Dunning}},\ }\href {\doibase 10.1063/1.456153} {\bibfield  {journal}
  {\bibinfo  {journal} {The Journal of Chemical Physics}\ }\textbf {\bibinfo
  {volume} {90}},\ \bibinfo {pages} {1007} (\bibinfo {year}
  {1989})}\BibitemShut {NoStop}%
\bibitem [{\citenamefont {Woon}\ and\ \citenamefont
  {Dunning~Jr}(1993)}]{woon1993gaussian}%
  \BibitemOpen
  \bibfield  {author} {\bibinfo {author} {\bibfnamefont {D.~E.}\ \bibnamefont
  {Woon}}\ and\ \bibinfo {author} {\bibfnamefont {T.~H.}\ \bibnamefont
  {Dunning~Jr}},\ }\href@noop {} {\bibfield  {journal} {\bibinfo  {journal} {J.
  Chem. Phys.}\ }\textbf {\bibinfo {volume} {98}},\ \bibinfo {pages} {1358}
  (\bibinfo {year} {1993})}\BibitemShut {NoStop}%
\bibitem [{\citenamefont {Olivares-Amaya}\ \emph {et~al.}(2015)\citenamefont
  {Olivares-Amaya}, \citenamefont {Hu}, \citenamefont {Nakatani}, \citenamefont
  {Sharma}, \citenamefont {Yang},\ and\ \citenamefont {Chan}}]{amaya2015}%
  \BibitemOpen
  \bibfield  {author} {\bibinfo {author} {\bibfnamefont {R.}~\bibnamefont
  {Olivares-Amaya}}, \bibinfo {author} {\bibfnamefont {W.}~\bibnamefont {Hu}},
  \bibinfo {author} {\bibfnamefont {N.}~\bibnamefont {Nakatani}}, \bibinfo
  {author} {\bibfnamefont {S.}~\bibnamefont {Sharma}}, \bibinfo {author}
  {\bibfnamefont {J.}~\bibnamefont {Yang}}, \ and\ \bibinfo {author}
  {\bibfnamefont {G.~K.-L.}\ \bibnamefont {Chan}},\ }\href
  {http://scitation.aip.org/content/aip/journal/jcp/142/3/10.1063/1.4905329}
  {\bibfield  {journal} {\bibinfo  {journal} {J. Chem. Phys.}\ }\textbf
  {\bibinfo {volume} {142}},\ \bibinfo {eid} {034102} (\bibinfo {year}
  {2015})}\BibitemShut {NoStop}%
\bibitem [{\citenamefont {Georges}\ \emph
  {et~al.}(1996{\natexlab{b}})\citenamefont {Georges}, \citenamefont {Kotliar},
  \citenamefont {Krauth},\ and\ \citenamefont {Rozenberg}}]{Kotliar1996}%
  \BibitemOpen
  \bibfield  {author} {\bibinfo {author} {\bibfnamefont {A.}~\bibnamefont
  {Georges}}, \bibinfo {author} {\bibfnamefont {G.}~\bibnamefont {Kotliar}},
  \bibinfo {author} {\bibfnamefont {W.}~\bibnamefont {Krauth}}, \ and\ \bibinfo
  {author} {\bibfnamefont {M.~J.}\ \bibnamefont {Rozenberg}},\ }\href@noop {}
  {\bibfield  {journal} {\bibinfo  {journal} {Rev. Mod. Phys.}\ }\textbf
  {\bibinfo {volume} {68}},\ \bibinfo {pages} {13} (\bibinfo {year}
  {1996}{\natexlab{b}})}\BibitemShut {NoStop}%
\bibitem [{\citenamefont {Holmes}\ \emph {et~al.}(2016)\citenamefont {Holmes},
  \citenamefont {Tubman},\ and\ \citenamefont {Umrigar}}]{holmes2016}%
  \BibitemOpen
  \bibfield  {author} {\bibinfo {author} {\bibfnamefont {A.~A.}\ \bibnamefont
  {Holmes}}, \bibinfo {author} {\bibfnamefont {N.~M.}\ \bibnamefont {Tubman}},
  \ and\ \bibinfo {author} {\bibfnamefont {C.~J.}\ \bibnamefont {Umrigar}},\
  }\href {\doibase 10.1021/acs.jctc.6b00407} {\bibfield  {journal} {\bibinfo
  {journal} {J. Chem. Theory Comp.}\ }\textbf {\bibinfo {volume} {12}},\
  \bibinfo {pages} {3674} (\bibinfo {year} {2016})},\ \Eprint
  {http://arxiv.org/abs/http://dx.doi.org/10.1021/acs.jctc.6b00407}
  {http://dx.doi.org/10.1021/acs.jctc.6b00407} \BibitemShut {NoStop}%
\bibitem [{\citenamefont {Loos}\ \emph {et~al.}(2018)\citenamefont {Loos},
  \citenamefont {Scemama}, \citenamefont {Blondel}, \citenamefont {Garniron},
  \citenamefont {Caffarel},\ and\ \citenamefont
  {Jacquemin}}]{loos2018mountaineering}%
  \BibitemOpen
  \bibfield  {author} {\bibinfo {author} {\bibfnamefont {P.-F.}\ \bibnamefont
  {Loos}}, \bibinfo {author} {\bibfnamefont {A.}~\bibnamefont {Scemama}},
  \bibinfo {author} {\bibfnamefont {A.}~\bibnamefont {Blondel}}, \bibinfo
  {author} {\bibfnamefont {Y.}~\bibnamefont {Garniron}}, \bibinfo {author}
  {\bibfnamefont {M.}~\bibnamefont {Caffarel}}, \ and\ \bibinfo {author}
  {\bibfnamefont {D.}~\bibnamefont {Jacquemin}},\ }\href@noop {} {\bibfield
  {journal} {\bibinfo  {journal} {J. Chem. Theory Comput.}\ } (\bibinfo {year}
  {2018})}\BibitemShut {NoStop}%
\end{thebibliography}

%

\setcounter{figure}{0}  
\setcounter{equation}{0}
\setcounter{table}{0}
\renewcommand{\thetable}{S\arabic{table}}   
\renewcommand{\thefigure}{S\arabic{figure}}

\section*{Supporting Information}
A development version of Q-Chem 5 was used for all calculations~\cite{shao_advances_2015}. 
\subsection*{G1 molecule set}

The G1 set~\cite{pople1989} is a set of small molecules that are useful for benchmark comparisons between theory and experiment.    Traditional coupled cluster theory or density functional theory with high level functionals perform well for most of these molecules, as demonstrated in various studies~\cite{tubman2018-1,tubman2018-2,feller1,mardirossian2017thirty,hait2018accurate}. However, these molecules have too many electrons for exact diagonalization through full configuration interaction with sufficiently large basis sets, and so the G1 set provides a valuable tool for exploring the usefulness of quantum algorithms for many-body systems.

\begin{figure*}
\begin{center}
\scalebox{1}{\includegraphics[width=2.2\columnwidth]{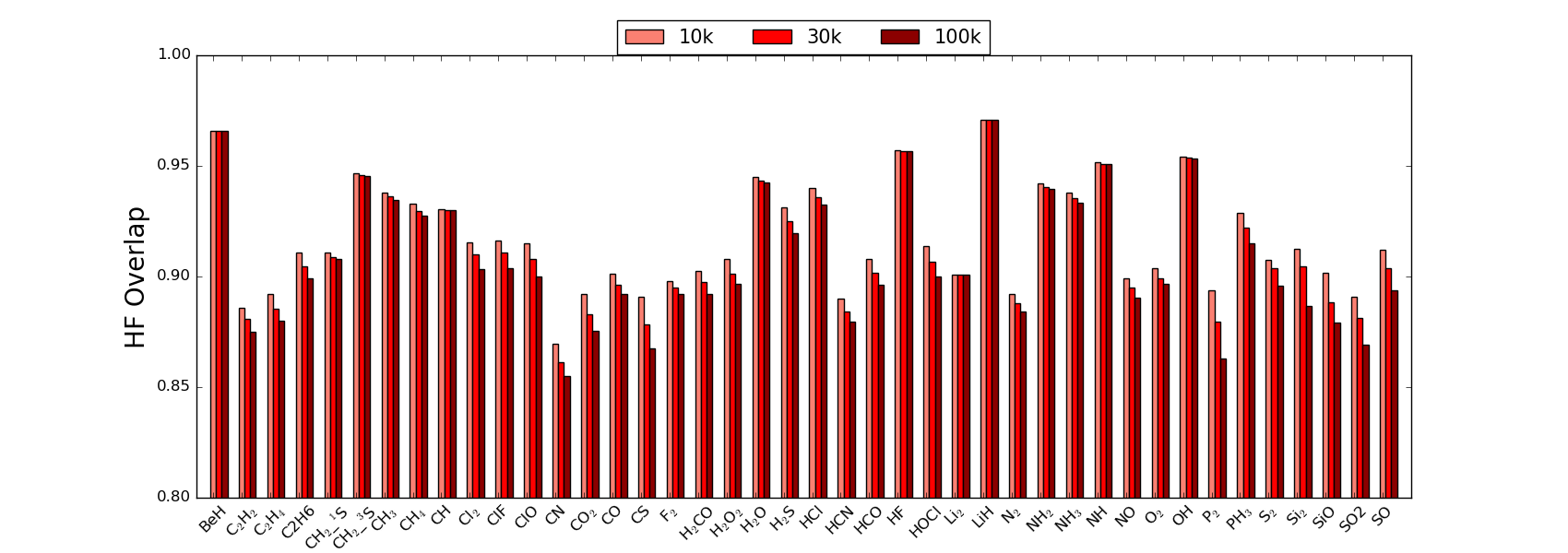}}
\end{center}
\caption{Histogram of overlap for the different molecules in the G1 set. For each molecule, we report results for ASCI calculations with $10^4$, $3\times 10^4$ and $10^5$ determinants using the cc-pVDZ basis set plus orbital rotation. See text for details. The denominators $^1S$ and $^3S$ for the methyl radical correspond to the singlet and triplet spin state.}
\label{fig:G1dets}
\end{figure*}

\begin{figure*}
\begin{center}
\scalebox{1}{\includegraphics[width=2.2\columnwidth]{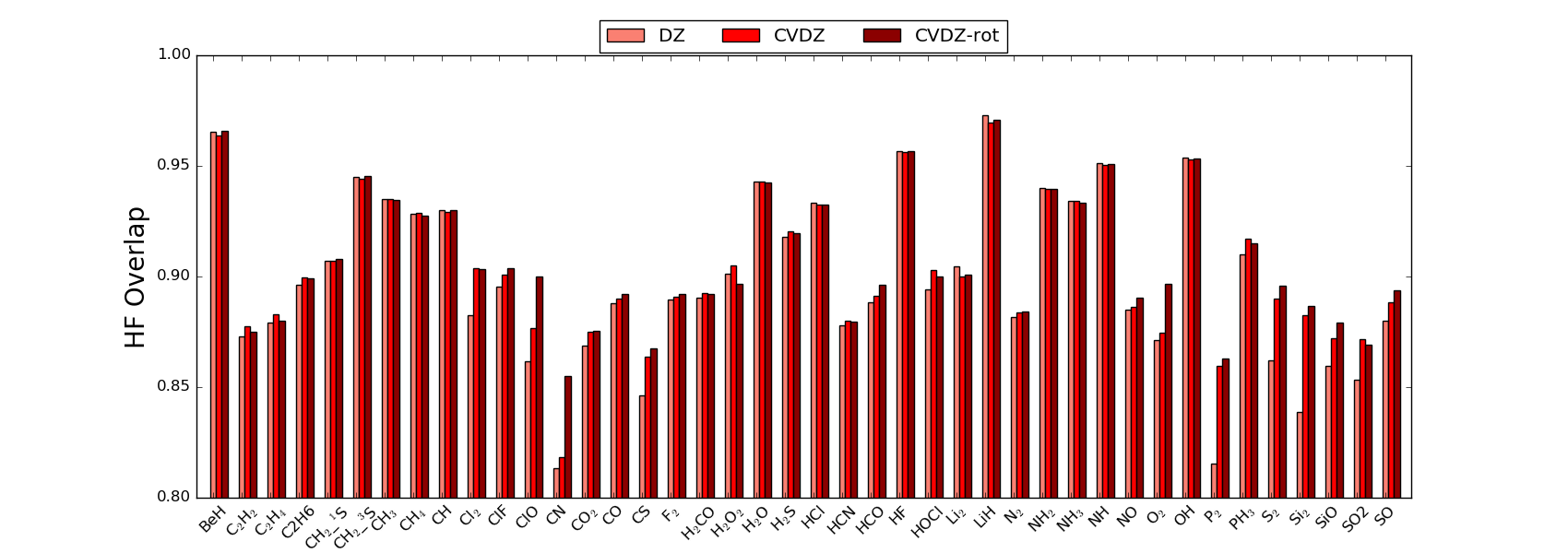}}
\end{center}
\caption{Histogram of overlap for the different molecules in the G1 set. For each molecule, we report results for ASCI calculations with $10^5$ determinants and different double zeta basis sets: DZ, cc-pVDZ (CVDZ) and cc-pVDZ plus orbital rotations (CVDZ-rot). See text for details. The denominators $^1S$ and $^3S$ for the methyl radical correspond to the singlet and triplet spin state.}
\label{fig:G1basis}
\end{figure*}

We performed ASCI calculations on the G1 molecule set. We studied the effect of using different basis sets and number of determinants in the ASCI calculation. For a subset of the molecules, Fig. \ref{fig:G1dets} and \ref{fig:G1basis} present the results.

We performed calculations with $10^4$, $3\cdot 10^4$ and $10^5$ determinants. Our calculations were done using a double zeta (DZ) basis set, a double zeta correlation-consistent (cc-pVDZ) basis set and the cc-pVDZ basis set with orbital rotation. The equivalent triple zeta (TZ) basis sets were also used, with little change of the overlaps as seen in the inset in Fig. \ref{fig:TZvsDZ} in the main paper. The cc-pVTZ~\cite{Dunning1989,woon1993gaussian} basis is designed for effective calculation of correlation energies (i.e. is correlation consistent) and has three basis functions for each valence energy level of constituent atoms (i.e. is triple zeta valence), along with extra polarization functions to distort the atomic densities for bonded environments. Such triple zeta polarized basis sets are known to yield good quality relative energies for standard mean-field approaches like Hartree-Fock and DFT, which is not the case for smaller basis sets~\cite{mardirossian2017thirty}.  The cc-pVTZ basis is often too large for most many body methods, and consequently only DRMG and selected CI methods, such as ASCI, have been able to achieve benchmark quality accuracy thus far~\cite{amaya2015,tubman2016-1,tubman2018-1,tubman2018-2}. 

\begin{figure}[htb!]
\begin{center}
\scalebox{1}{\includegraphics[width=\columnwidth]{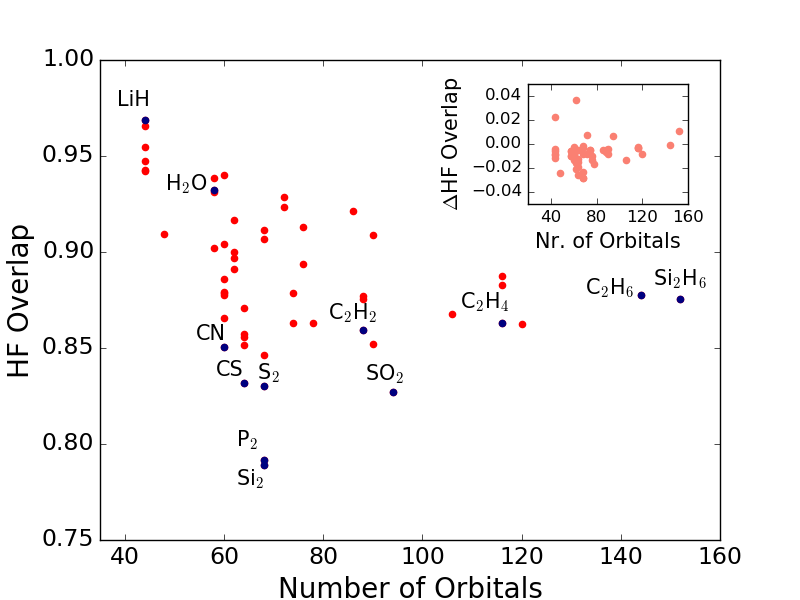}}
\end{center}
\caption{Scatter plot of systems with respect to number of orbitals for cc-pVTZ basis set.  Labeled calculations are highlighted in dark blue.  The inset shows the difference between the Hartree-Fock overlap in the cc-pVDZ and cc-pVTZ basis. See text for more explanation.}
\label{fig:TZvsDZorb}
\end{figure}

\begin{figure}[htb!]
\begin{center}
\scalebox{1}{\includegraphics[width=\columnwidth]{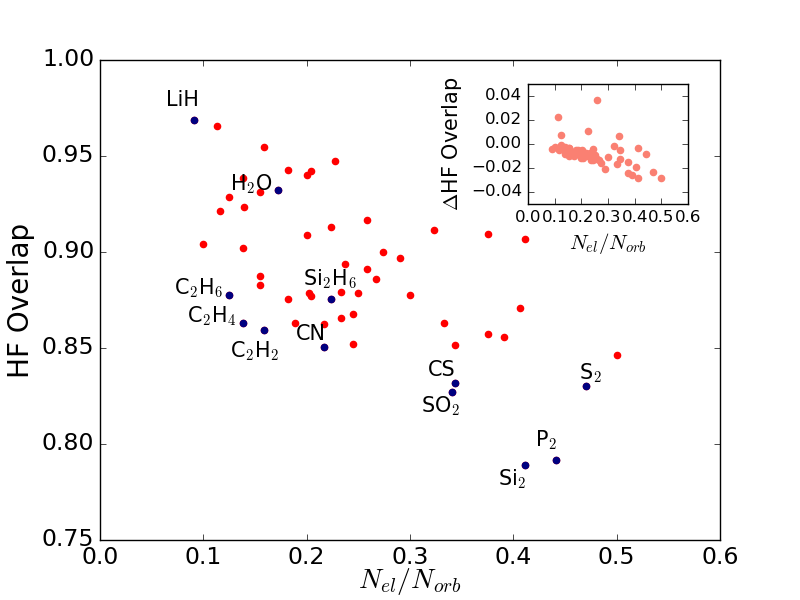}}
\end{center}
\caption{Scatter plot of systems with respect to the filling fraction for cc-pVTZ basis set.  Labeled calculations are highlighted in dark blue.  The inset shows the difference between the Hartree-Fock overlap in the cc-pVDZ and cc-pVTZ basis. See text for more explanation.}
\label{fig:TZvsDZfill}
\end{figure}

In Fig. \ref{fig:TZvsDZ} we report the overlaps as a function of the number of electrons. Other possible representations, namely as a function of orbitals and of the filling fraction (i.e. number of electrons divided by the number of orbitals), are shown in Fig. \ref{fig:TZvsDZorb} and \ref{fig:TZvsDZfill}. While the trends in these three plots are very similar, the spread of points is minimized when plotting the overlap as a function of the number of electrons in Fig. \ref{fig:TZvsDZ}.

\subsection*{Hubbard Model Calculations}

We performed the two dimensional square lattice Hubbard model calculations on finite square lattice clusters with periodic boundary conditions. We used two basis sets: The spatial orbital basis set that is often used to introduce the Hubbard model Hamiltonian
\begin{equation}
H = -t\sum_{\langle i,j \rangle,\sigma}\left(c_{i\sigma}^\dagger c_{j\sigma} + h.c.\right) + U\sum_i n_{i\uparrow}n_{i\downarrow},
\label{Eq:HubAO}
\end{equation} 
where $c_{i\sigma}^\dagger$ is the creation operator for a Fermion of spin $\sigma$ on lattice site $i$, $n_{i\sigma}=c_{i\sigma}^\dagger c_{i\sigma}$, $t$ is the hopping amplitude between nearest neighbors and $U$ is the local Coulomb repulsion. The other basis used is a plane wave basis, defined from the spatial basis as

\begin{equation}
\tilde{c}^\dagger_{k\sigma} = \frac{1}{\sqrt{N}}\sum_{j}e^{-i\ j\cdot k}c^\dagger_{j\sigma},
\end{equation}
where $k$ is one of the allowed lattice momenta and $N$ the total number of lattice sites. The value of the allowed lattice momenta are determined by the size of the lattice cluster and the periodic boundary conditions. In this basis, the Hamiltonian takes the following form
\begin{equation}
\begin{split}
H = &-2t\sum_{k,\sigma}(\cos{k_x} + \cos{k_y})\tilde{c}_{k\sigma}^\dagger \tilde{c}_{k\sigma}\\ 
&+ \frac{U}{N}\sum_{k,p,q} \tilde{c}_{k+q\uparrow}^\dagger \tilde{c}_{k\uparrow}\tilde{c}_{p-q\downarrow}^\dagger \tilde{c}_{p\downarrow},
\label{Eq:HubPW}
\end{split}
\end{equation} 
where $k_x$ and $k_y$ are the x and y components of the momentum $k$. From Eq. \ref{Eq:HubPW}, it can been seen that the interaction term conserves total momentum, and thus the Hilbert space splits into sectors characterized by the total momentum that are not connected through the Hamiltonian.

For the calculations presented in Fig. \ref{fig:HubRes}, square lattice clusters with 16 (4x4), 25 (5x5) and 36 (6x6) sites were simulated. The Coulomb repulsion was calculated with $U / t = 1, 4 $ and 8.  Different electron densities were simulated by changing the number of electrons in the simulation. The electron fillings given correspond to the number of electrons per site divided by two, so that one electron per site is reported as $50\%$ filling (half-filling). All simulations, except the half-filling (5x5) clusters, had the same numbers of electrons with spin up and spin down. The simulation parameters are summarized in Tab. \ref{tab:HubParams}.

\begin{table*}[htb!]
\centering
\begin{tabular}{|c|c|c|c|c|c|}
\hline
&$12.5\%$&$20\%$&$25\%$& $40\%$ & $50\%$  \\
\hline
(4x4)&($2\uparrow,2\downarrow$), 912 dets (ED)&($3\uparrow,3\downarrow$), $1.96\cdot 10^4$ dets (ED)& ($4\uparrow,4\downarrow$), $2.07184\cdot 10^5$ dets (ED)& ($6\uparrow,6\downarrow$), $10^6$ dets&($8\uparrow,8\downarrow$), $10^6$ dets\\
(5x5)&($3\uparrow,3\downarrow$), $2.116\cdot 10^5$ dets (ED)&($5\uparrow,5\downarrow$), $10^6$ dets& ($6\uparrow,6\downarrow$), $10^6$ dets& ($10\uparrow,10\downarrow$), $10^6$ dets&($13\uparrow,12\downarrow$), $10^6$ dets \\
(6x6)&($5\uparrow,5\downarrow$), $10^7$ dets&($7\uparrow,7\downarrow$), $10^7$ dets& ($9\uparrow,9\downarrow$), $10^7$ dets& ($14\uparrow,14\downarrow$), $10^7$ dets&($18\uparrow,18\downarrow$), $10^7$ dets \\
\hline
(4x4)&($2\uparrow,2\downarrow$), $1.44\cdot 10^4$ dets (ED)&($3\uparrow,3\downarrow$), $10^5$ dets& ($4\uparrow,4\downarrow$), $10^5$ dets& ($6\uparrow,6\downarrow$), $10^5$ dets&($8\uparrow,8\downarrow$), $10^5$ dets \\
(5x5)&($3\uparrow,3\downarrow$), $10^5$ dets &($5\uparrow,5\downarrow$), $10^5$ dets & ($6\uparrow,6\downarrow$), $10^5$ dets & ($10\uparrow,10\downarrow$), $10^5$ dets &($13\uparrow,12\downarrow$), $10^5$ dets  \\
(6x6)&($5\uparrow,5\downarrow$), $10^5$ dets &($7\uparrow,7\downarrow$), $10^5$ dets & ($9\uparrow,9\downarrow$), $10^5$ dets & ($14\uparrow,14\downarrow$), $10^5$ dets &($18\uparrow,18\downarrow$), $10^5$ dets  \\
\hline
\end{tabular} 
\caption{Number of electrons and determinants in the ASCI simulations of the two dimensional square lattice Hubbard model presented in Fig. \ref{fig:HubRes} in the main body of the paper. The rows correspond to cluster sizes, the columns to electron fillings. The first three rows refer to the plane wave calculations, the last three to the spatial basis. The calculations where the full Hilbert space was included are marked with (ED), meaning exact diagonalization. For each cluster size and electron filling we performed three simulations with different Coulomb repulsions $U/t = 1, 4$ and 8.}
\label{tab:HubParams}
\end{table*}%

The Hubbard Hamiltonian in the spatial basis is extremely sparse, states in the Hilbert space are exclusively connected by single excitations related to electron hops between nearest neighbors. At half-filling and high $U/t$, the two dimensional square lattice Hubbard model is expected to develop an anti-ferromagnetic Mott insulating phase, that can be described mainly with one single determinant in the spatial basis (two in the case of equal number of spin up and spin down electrons). Away from half-filling, the Hubbard model presents different phases, including non-homogeneous charge density orders \cite{Zheng2017}. In order to target a particular phase, we start the ASCI calculations with a determinant presenting the order intended. For example, we used the anti-ferromagnetic state as initial determinant in the ASCI method in the spatial basis at half-filling. 
As a simple comparison of the using starting determinants with different orders, for the (6x6) cluster at half-filling and $U/t = 8$ the ASCI calculation with $10^6$ determinants starting with the anti-ferromagnetic state provided an energy per site of $-0.43\ t$ while an ASCI calculation with the same number of determinants starting from a spin-density wave ordered determinant provided a final energy of $-0.28\ t$. The overlaps of the first determinants with the total wavefunction were approximately $20\%$ and $6\%$ respectively. The higher energy and lower overlap in the calculation starting from a spin-density wave indicates that this is not the right ground state phase for the square lattice Hubbard model at half-filling and $U / t = 8$.

For the plane wave basis, proposing a single determinant with a particular kind of spatial order is less evident. Furthermore, the effect of the starting determinant is less dramatic due to (i) the decomposition of the full Hilbert space into sectors characterized by the total momentum and (ii) the higher connectivity inside these Hilbert subspaces through the Hamiltonian. We limit the ASCI search to the subspace of zero momentum states, where we expect the true ground state to be found.

\begin{figure}[htb!]
\begin{center}
\scalebox{1}{\includegraphics[width=\columnwidth]{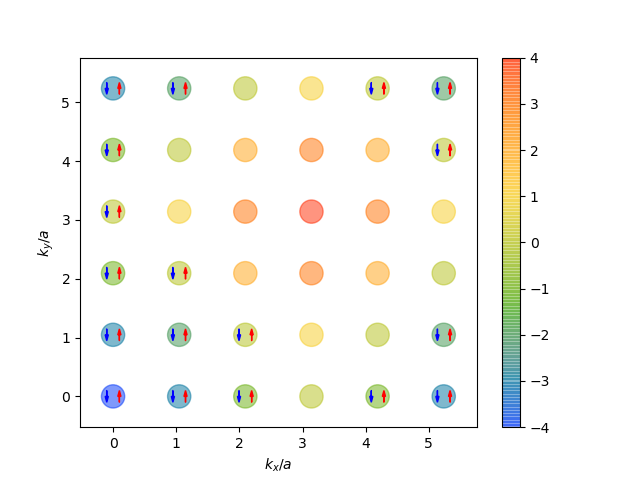}}
\end{center}
\caption{Fermi sea zero momentum state for the (6x6) cluster at half-filling in the plane wave space. Each circle corresponds to one momentum state, it's occupation is symbolized by arrows representing electrons with a particular spin state. Spin up is represented with a red arrow, spin down with a blue arrow. The color map corresponds to the lattice kinetic energy for each degree of freedom (plane wave state), as given by the dispersion given in Eq. \ref{Eq:HubPW}.}
\label{fig:init_state1}
\end{figure}

Another direct consequence of points (i) and (ii) above is that the ASCI search in plane-wave space is much more efficient than its spatial basis counterpart. The ground state energies per site for the plane wave simulations are thus far below the ground state energies per site for the spatial basis when using the same number of determinants, except for the case of half-filling at high $U/t$, where the true ground state has a single determinant representation in the spatial basis. Since ASCI is a variational method, the plane wave solutions are a more reliable representation of the true ground state of the system.

\begin{figure}[htb!]
\begin{center}
\scalebox{1}{\includegraphics[width=\columnwidth]{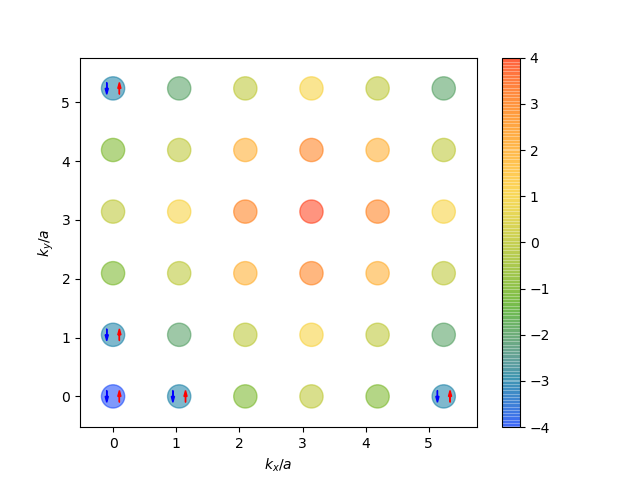}}
\end{center}
\caption{Most important zero momentum state for the (6x6) cluster at $12.5\%$ filling and $U/t = 1$ in the plane wave space.
See Fig. \ref{fig:init_state1} for details.
}
\label{fig:closedShell}
\end{figure}

\begin{figure*}
\begin{center}
\scalebox{1}{\includegraphics{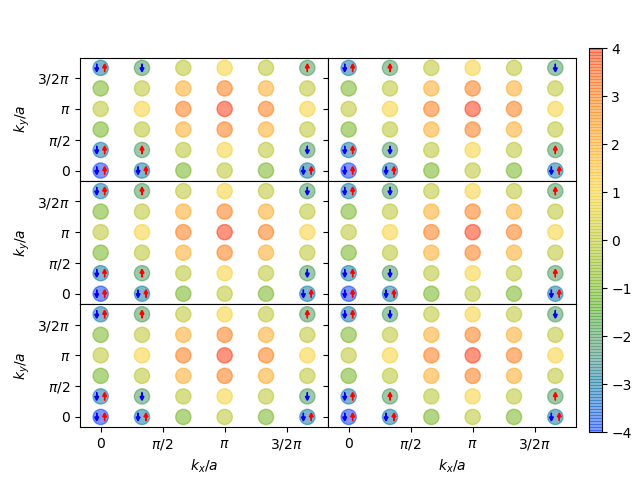}}
\end{center}
\caption{Six most important zero momentum states for the (6x6) cluster at $20\%$ filling and $U/t = 1$ in the plane wave space. Each circle corresponds to one momentum state, it's occupation is symbolized by arrows representing electrons with a particular spin state. Spin up is represented with a red arrow, spin down with a blue arrow. The color map corresponds to the kinetic energy of each degree of freedom as given in Eq. \ref{Eq:HubPW}.
}
\label{fig:openShell}
\end{figure*}

The overlaps shown in Fig. \ref{fig:HubRes}a can be understood  by looking at the underlying the shell structures that arise in the Hubbard model when using a plane wave basis. As an example we represent graphically in Fig. \ref{fig:init_state1} the occupation of the momentum degrees of freedom for the most important determinant of a (6x6) calculation at half filling.   Single particle states with the same energy (the same color in the heat map) form open shells. The states with overlap close to 1 in Fig. \ref{fig:HubRes}a correspond to densities with an effectively closed shell.   A closed shell is shown in Fig. \ref{fig:closedShell} for a (6x6) Hubbard model at $12.5\%$ filling.   As U/t increases the shell structure becomes less important. However, for U/t=8, our results show that the influenced of the shell structure is still quite large.   This is key to understand the quantum states with low single determinant overlaps, as they correspond to effects from the open shells, see Fig. \ref{fig:openShell}. For these cases, there is a group of several determinants with degenerate absolute coefficient in the full many-body wavefunction. For the case of the (6x6) at $20\%$  filling, the sum of the coefficient square of these top determinants corresponds to about $97\%$ of the total wave function. Thus, for all systems where the plane-wave basis performs well (i.e. for all parameters explored in the paper except for half-filling at high $U/t$), a small number of determinants sum up to approximately $90\%$ of the total wave function.

\subsection*{Embedding Methods}

The overlaps for Hubbard-Anderson Hamiltonians presented in Fig. \ref{fig:Embed} of the paper were computed using the ASCI-DMFT embedding method. Dynamical mean-field theory (DMFT), and it's cluster version used here, is a non-perturbative method for the treatment of strongly correlated many-body systems \cite{Kotliar1996}. Usually applied to lattice models, the main idea is to identify a sub-lattice, which we will call the cluster, where all interactions are treated exactly. The coupling of the cluster to the rest of the system is then represented by a set of non-interacting bath sites which couple to all sites inside the cluster independently. The parameters defining those bath sites, namely the bath single-particle energies and the coupling terms between bath and cluster sites, are determined self-consistently so that the Green's function of the cluster+bath system, also called impurity model, coincides with the local Green's function of the original lattice when limited to the cluster sites. For zero temperature calculations, the ground state of the impurity model has to be computed in each iteration. A wide range of approaches is currently used to compute this ground state, and determining an optimal solver is an currently active field of research. In this paper, we chose to represent configuration interaction based solvers by using the ASCI algorithm as DMFT solver \cite{Mejuto2017}.   

The reliability of DMFT calculations depends on how many degrees of freedom are included in the cluster and for a subfamily of DMFT methods it also depends on the number of bath sites. Thus, we performed ASCI-DMFT calculations for the one band two dimensional square lattice Hubbard model, see Eq. \ref{Eq:HubAO}, with cluster sizes (2x2), (3x3) and (4x4) and different number of baths for different particle fillings. These should exemplify a wide range of complexities of configuration interaction based DMFT calculations for the two dimensional Hubbard model. Since DMFT is a self-consistent algorithm, the impurity model is solved multiple times for given cluster and bath sizes. For each one of these intermediate solutions, we computed the overlap of the single most important determinant in the full ASCI ground state wave function. Then, for each combination of cluster size, bath size and particle filling, we computed the averaged overlap over all iterations in the DMFT procedure. All DMFT calculations were converged within 10 to 20 iterations.

\subsection*{Homogenous Electron Gas}
The overlaps in  Table \ref{tab:ueg} of the paper were computed between the ASCI wavefunction and the aufbau filled slater determinant in the plane wave basis (i.e. the determinant where the orbitals are filled in ascending order of energy). This determinant is a solution to the Hartree-Fock equations, and is therefore described as the Fermi gas Hartree-Fock solution in the paper.

The squared overlaps were linearly extrapolated against the second order perturbation theory (PT2) correction to the ASCI variational energy. The PT2 energy is a measure of the incompleteness of the ASCI wavefunction, and linear extrapolation of selected CI energies against the corresponding PT2 corrections is already widely used in literature\cite{holmes2016,loos2018mountaineering}. The
limit of zero PT2 energy was assumed to be the actual overlap. The extrapolation however did not lead to any change in the second decimal place (relative to the estimate from best variational wavefunction) for $r_s\le 2$, and so we are reasonably confident that those numbers have fully converged. The more strongly correlated $r_s=5$ and $10$ however had a substantial difference ($>0.01$) between the overlap squared calculated from the best variational wavefunction and the extrapolated result. This difference represents an estimate of the error in the overlap, and is reported in Table \ref{tab:ueg}. 

\begin{figure}[htb!]
\begin{center}
\scalebox{1}{\includegraphics[width=\columnwidth]{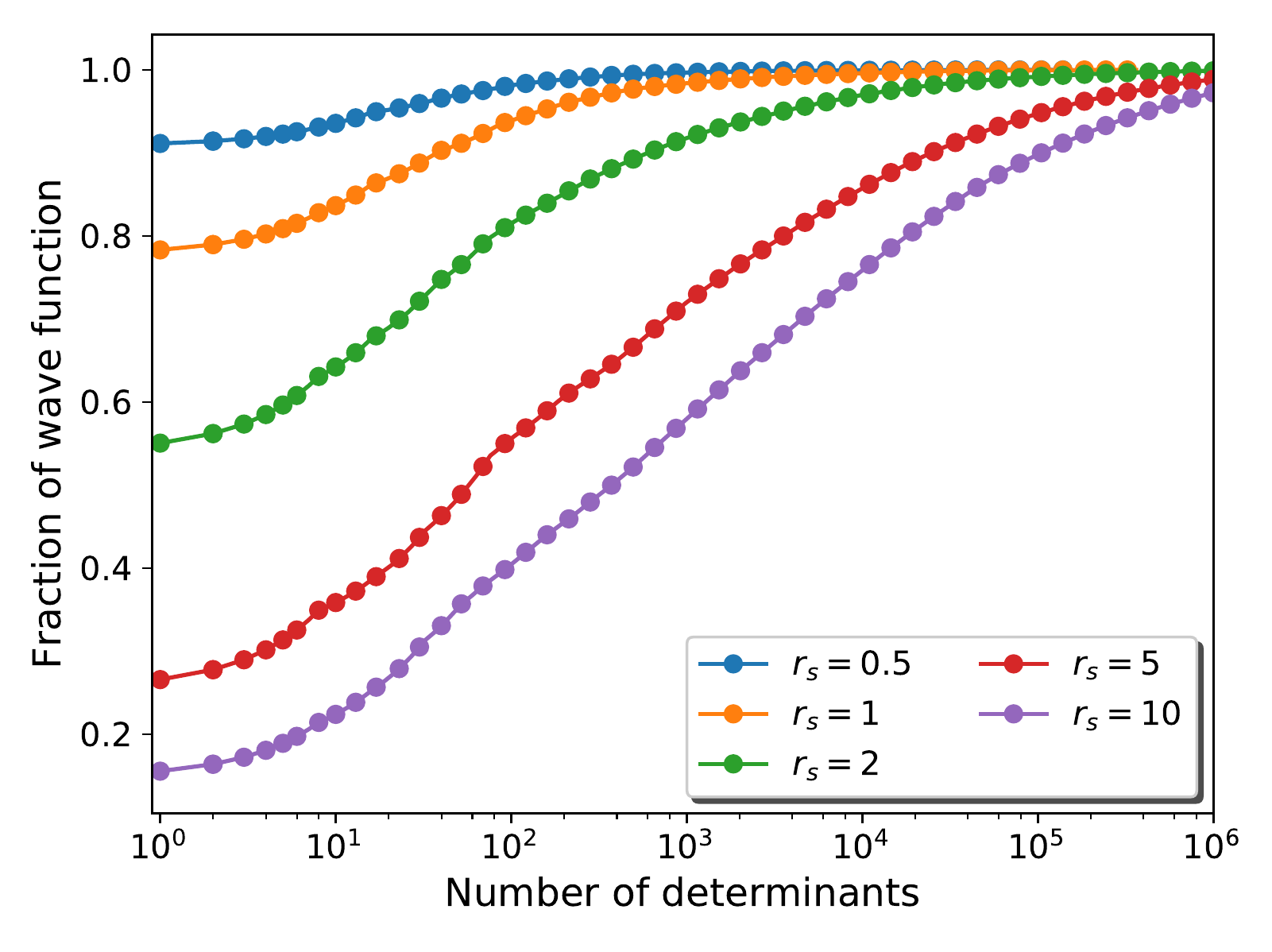}}
\end{center}
\caption{Fraction of the ASCI wave function constituted by the most important $N$  Slater determinants for 2D HEG (10 electrons in 69 orbitals), at various $r_s$.  Note that the $x$ axis is on a log scale.}
\label{fig:uegcum}
\end{figure}

In Fig. \ref{fig:uegcum} we show the rapid improvement of ground state overlap with number of determinants. We see that there is one determinant with very large overlap at small $r_s$, but this decreases quite a bit for large $r_s$ where the system is more strongly correlated. We however do observe that the first $10^2-10^3$ determinants appear to collectively make up more than half of the wave function even for large $r_s=5-10$. Multideterminant initial state preparation therefore could assist simulations of such strongly correlated systems on quantum devices. 

\end{document}